\documentclass[11pt]{article}
\usepackage{amsfonts}

\usepackage{amscd}
\usepackage{amsmath}
\newlength{\dinwidth}
\newlength{\dinmargin}
\def\be{\begin{equation}}
\def\ee{\end{equation}}
\def\ben{\begin{displaymath}}
\def\een{\end{displaymath}}
\def\baa{\begin{eqnarray}}
\def\eaa{\end{eqnarray}}

\def\ba{\begin{array}}
\def\ea{\end{array}}

\makeatletter
\@addtoreset{equation}{section}
\makeatother

\def\a{\alpha}
\def\g{\gamma}

\def\b{\beta}

\def\l{\lambda}

\def\O{\Omega}
\def\lp{{x_m}}

\def\CP1{{\mathbb P}^1}

\def\la{\label}

\def\f{\frac}
\def\L{{\cal L}}
\def\p{\partial}

\def\tr{{\rm tr}}
\def\log{\ln}

\def\la{\label}

\def\f{\frac}
\def\L{{\cal L}}
\def\p{\partial}
\def\res{{\rm res}}

\def\det{{\rm det}}

\def\Ld{\L_{{\rm dissected}}}
\newtheorem{remark}{Remark}
\newtheorem{definition}{Definition}
\newtheorem{theorem}{Theorem}
\newtheorem{corollary}{Corollary}
\newtheorem{lemma}{Lemma}

\begin{document}

\title{Tau-functions on  Hurwitz spaces}
\author{A. Kokotov\footnote{e-mail: alexey@mathstat.concordia.ca, phone (1) (514) 8483246}   and
D. Korotkin
\footnote{e-mail: korotkin@mathstat.concordia.ca, phone:(1)(514)8483245, fax: (1)(514)8482831}}
\maketitle

\begin{center}
Department of Mathematics and Statistics, Concordia University\\
7141 Sherbrook West, Montreal H4B 1R6, Quebec,  Canada
\end{center}


\vskip0.5cm
{\bf Abstract.}
We construct a flat holomorphic line bundle over a connected
component of the Hurwitz space of branched coverings of the Riemann sphere ${\mathbb P}^1$.
A flat holomorphic connection defining the bundle is described in terms of the invariant Wirtinger projective
connection on the branched covering corresponding to a given meromorphic function on a Riemann surface of genus
$g$. In genera 0 and 1 we construct a nowhere vanishing holomorphic
horizontal section of this bundle (the ``Wirtinger tau-function''). In
higher genus we compute the modulus square of the Wirtinger
tau-function.
\newpage
\vskip0.5cm
\section{Introduction}
Holomorphic line bundles over moduli spaces of Riemann surfaces were studied by many researchers
during last 20 years (see, e. g., Fay's survey \cite{Fay}).
In the present paper we consider (flat) holomorphic line bundles over Hurwitz spaces
(the spaces of meromorphic functions on Riemann surfaces or, what is the same,
the spaces of branched coverings of the Riemann sphere ${\mathbb P}^1$) and over
coverings of Hurwitz spaces. The covariant constant
sections  (we call them tau-functions) of these bundles are the main object of our consideration.

Our work was inspired by a coincidence of the isomonodromic tau-function of a class of $2\times 2$
Riemann-Hilbert problems solved in \cite{KitKor} with the heuristic
expression which appeared in the context of the string theory and was interpreted as the determinant
of the Cauchy-Riemann operator acting in a spinor line bundle over a hyperelliptic Riemann
surface (see the survey \cite{Knizhnik}).

To illustrate our results consider, for example, the Hurwitz space $H_{g, N}(1, \dots, 1)$
consisting  of $N$-fold coverings of genus $g$
with only simple branch points, none of which coincides with infinity.
(In the main text we work with coverings having branch points of arbitrary order.)

Let $\L$ be a covering from $H_{g, N}(1, \dots, 1)$,
we use the branch points $\l_1, \dots, \l_M$ (i. e. the projections of the ramification points $P_1,
\dots, P_M$ of the covering $\L$)  as local coordinates on
the space $H_{g, N}(1, \dots, 1)$; according to the Riemann-Hurwitz formula $M=2g+2N-2$.

Let $\l$ be the coordinate of the projection of a point $P\in \L$ to ${\mathbb P}^1$. In a neighborhood of a
ramification point $P_m$ we introduce the
local coordinate $x_m=\sqrt{\l-\l_m}$.

Besides the Hurwitz space $H_{g, N}(1, \dots, 1)$, we shall use the
``punctured'' Hurwitz space ${H}_{g, N}'(1, \dots, 1)$, which is
obtained from $H_{g, N}(1, \dots, 1)$ by excluding all branched
coverings  which have at least one vanishing theta-constant.

In the trivial bundle ${H}_{g, N}'(1, \dots, 1)\times {\mathbb C}$ we introduce
the connection
\begin{equation}\label{connect}
d_W=d-\sum_{m=1}^M{\cal A}_md\l_m\;,
\end{equation}
where $d$ is the external differentiation operator including both
holomorphic and antiholomorphic parts; connection coefficients are
expressed in terms of the invariant Wirtinger projective connection
$S_W$ on the covering $\L$
as follows:
\begin{equation}\label{akrasiv}
{\cal A}_m=-\frac{1}{12}S_W(x_m)\Big|_{x_m=0}, \ \ \ m=1, \dots M\,.
\end{equation}
The connection coefficients ${\cal A}_m$ are holomorphic with respect
to $\l_m$ and  well-defined for all
coverings $\L$ from the ``punctured'' Hurwitz space $H'_{g, N}(1, \dots, 1)$.

Connection (\ref{connect}) turns out to be flat; therefore, it
determines a character of the fundamental group of
${H}_{g, N}'(1,\dots, 1)$; this character defines a flat holomorphic line bundle ${\cal T}_W$
over ${H}_{g, N}'(1,\dots, 1)$. We call this bundle the ``Wirtinger
line bundle'' over Hurwitz space; its horizontal  holomorphic section we
call the Wirtinger tau-function of the covering $\L$.

In a trivial bundle $U(\L_0)\times {\mathbb C}$, where $U(\L_0)$ is a small
neighborhood of a given covering $\L_0$ in $H_{g, N}(1, \dots , 1)$ we can define also the flat
connection $d_B=d-\sum_{m=1}^M {\cal B}_m d\l_m$, where the coefficients
${\cal B}_m$
are built from the  Bergmann projective connection $S_B$ in a  way similar
to (\ref{akrasiv}):
$${\cal B}_m=-\f{1}{12} S_B(x_m)\Big|_{x_m=0}\;.$$
The  covariant constant section of this line bundle
in case of hyperelliptic coverings ($N=2$, $g>1$)
turns out to coincide (see
\cite{KitKor} for explicit calculation) with heuristic expression for the determinant of the Cauchy-Riemann
operator acting in the trivial line bundle over a hyperelliptic
Riemann surface, which was proposed in \cite{Knizhnik}. This  section also
appears as a part of isomonodromic tau-function associated to matrix
Riemann-Hilbert problems with quasi-permutation monodromies \cite{Koro01}.
However, since the Bergmann projective connection, in contrast to
Wirtinger projective connection, {\it does} depend on the choice of
canonical basis of cycles on the covering, connection $d_B$ can not be
globally continued to the whole Hurwitz space, but
only to its appropriate covering. We call the corresponding line bundle over this covering
the Bergmann line bundle and its covariant constant section -- the Bergmann tau-function.

We obtain explicit formulas for the modulus square of the Wirtinger and Bergmann
 tau-functions in
genus greater than $1$; in genera $0$ and $1$ we perform the
``holomorphic factorization'' and derive explicit formulas for the
tau-functions themselves.

In genera  1 and 2 (as well as in genus $0$) there are no vanishing theta-constants,
i.e.  $H_{g, N}(1, \dots, 1)= H_{g, N}'(1, \dots, 1)$;  therefore,
the holomorphic bundle
${\cal T}_W$ is the bundle over the whole Hurwitz space $H_{g, N}(1, \dots, 1)$.

To write down  an explicit formula for the tau-function
 over the Hurwitz space $H_{1, N}(1, \dots, 1)$,
consider a holomorphic (not necessarily normalized) differential
 $v(P)$  on an elliptic covering
$\L\in H_{1, N}(1, \dots, 1)$.
Introduce the notation $f_m\equiv f_m(0)$, $h_k\equiv h_k(0)$, where
$v(P)=f_m(x_m)dx_m$ near the branch point $P_m$
and $v(P)=h_k(\zeta)d\zeta$ near
the infinity of the $k$-th sheet; $\zeta=1/\lambda$, where $\l$ is the coordinate of the
projection of a point
$P\in\L$ to ${\mathbb P}^1$.
Then the Wirtinger tau-function on  $H_{1, N}(1, \dots, 1)$ is given by the formula
\begin{equation}\label{otvrod11}
\tau_W=\frac{\left\{\prod_{k=1}^N h_k\right\}^{1/6}}{\left\{\prod_{m=1}^M f_m\right\}^{1/12}}\;.
\end{equation}
\vskip0.5cm
The analogous explicit formula can be written for coverings of genus $0$.

The results in genera $0, 1$ follow from the study of
the properly regularized Dirichlet integral ${\mathbb S}=
\frac{1}{2\pi}\int_{\L}|\phi_\l|^2$, where $e^\phi |d\l|^2$ is
the flat metric on $\L$ obtained by projecting down the
standard metric $|dz|^2$ on the  universal covering
$\tilde\L$. The derivatives of ${\mathbb S}$ with respect to the
branch points can be expressed through the values of the
Schwarzian connection at the branch points;
 this reveals a close link of ${\mathbb S}$ with the modulus of
the tau-function. On the other hand, the integral ${\mathbb S}$ admits
an explicit calculation via the asymptotics of the flat metric
near the branch points and the infinities of the sheets of the
covering. Moreover, it admits a ``holomorphic factorization''
i.e. it can be explicitly represented as the modulus square of some
holomorphic function, which allows one to compute the tau-function
itself.

The same tools (except the explicit holomorphic factorization)
also work in case of higher genus, when two equivalent approaches are possible.

First, one can exploit the Schottky uniformization and introduce the Dirichlet integral
corresponding to the flat metric  on $\L$ obtained by projecting of the flat metric $|d\omega|^2$
on a fundamental domain of the Schottky group. This approach leads to the expression of the
modulus square of the tau-function through the holomorphic function $F$
on the Schottky space,
which was introduced in \cite{Zograf} and can be interpreted as the holomorphic determinant
of the Cauchy-Riemann operator acting in the trivial line bundle over
$\L$.
(In the main text we denote this function directly by ${\rm det}\,\bar\partial$.)

The second approach uses the Fuchsian uniformization and the Liouville action corresponding to the
metric of constant curvature $-1$ on $\L$. It gives the following expression
for the modulus square of the tau-function:
\begin{equation}\label{dettau}
|\tau_W|^2=e^{-{\mathbb S}_{Fuchs}/6}\frac{{\rm det}\Delta}{{\rm det}\,\Im\,{\mathbb B}}
\prod_{\beta\,{\rm  even}}\Big|\Theta[\beta](0\,|\,{\mathbb B})
\Big|^{-\frac{8}{4^g+2^g}},
\end{equation}
where ${\rm det}\Delta$ is the determinant of the Laplacian on the $\L$; ${\mathbb S}_{Fuchs}$ is
an appropriately regularized Liouville action which is a real-valued function of the
 branch points; ${\mathbb B}$ is the matrix of
$b$-periods of the branched covering.

Existence of explicit holomorphic factorization of our expressions
for $|\tau_W|^2$ in genera $g=0, 1$  allows to suggest that explicit formulas
for $\tau_W$ similar to (\ref{otvrod11}) also exist in higher genera.

In this paper we use the technical tools developed
in \cite{ZT2, ZT3}. We strongly suspect that in our context
it should be possible to avoid the extrinsic formalism of the Dirichlet
integrals and Liouville action and, at the least,
it should exist a direct way to prove the genus 1 formula (\ref{otvrod11}).

The paper is organized as follows.
In section 2 after some preliminaries
we prove the flatness of the  connections $d_W$ and $d_B$ and introduce the
flat line bundles over Hurwitz spaces and their coverings. In section 2 we
 find explicitly the tau-functions for genera $0$ and $1$.
 In section 3, using the Schottky and Fuchsian uniformizations, we give
the expressions for  the
modulus square of tau-functions in genus greater than $1$.

Our work on this paper was greatly influenced by  Andrej Nikolaevich
Tyurin; in particular,
he attracted our attention to the
Wirtinger bidifferential.

The authors
are also greatly indebted to the anonymous referee; a lot
of his proposals and remarks were used here.

This work was partially supported by the grant of
Fonds pour la Formation de Chercheurs et l'Aide a la Recherche de
Quebec, the grant of Natural Sciences and Engineering Research Council
of Canada and Faculty Research Development Program of Concordia University.

\section{Tau-functions of branched coverings}
\subsection{The Hurwitz spaces}
Let $\L$ be a compact Riemann surface of genus $g$ represented as an $N$-fold
branched covering
\begin{equation}\label{nakr}
p: \L \longrightarrow {\mathbb P}^1,
\end{equation}
of the Riemann sphere ${\mathbb P}^1$.
Let the holomorphic map $p$ be ramified at the points $P_1, P_2, \dots, P_M\in \L$ of ramification
indices  $r_1, r_2, \dots, r_M$ respectively (the ramification index is equal to the number of sheets
glued at a given ramification point).
 Let also $\l_m=p(P_m), \ m=1,2, \dots, M$ be the branch points.
(Following \cite{Fulton},
we reserve the name "ramification points" for the points $P_m$ of the surface $\L$ and the name
"branch points" for the points $\l_m$ of the base ${\mathbb P}^1$.)

  We assume that none of the branch points $\l_m$  coincides
with the infinity and  $\l_m\neq\l_n$ for $m\neq n$.

Recall that two branched coverings $p_1:\L_1\rightarrow {\mathbb P}^1$ and
$p_2:\L_2\rightarrow {\mathbb P}^1$ are called equivalent
if there exists a biholomorphic map $f:\L_1\rightarrow \L_2$ such that $p_2f=p_1$.
Let $H(N, M, {\mathbb P}^1)$ be the Hurwitz space of the equivalence classes of $N$-fold branched coverings
of ${\mathbb P}^1$ with $M$ branch points none of which coincides with the infinity.
 This space can be equipped with natural topology (see \cite{Fulton})
and is a (generally disconnected) complex manifold. Denote by ${\cal U}(\L)$ the connected component of
$H(N, M, {\mathbb P}^1)$ containing
the equivalence class of the covering $\L$.
According to the Riemann-Hurwitz formula, we have
$$g=\sum_{m=1}^M\frac{r_m-1}{2}-N+1,$$
where $g$ is the genus of the surface $\L$.

   If all the branch points of the covering $\L$ are simple (i. e. all the $r_m$ are equal to $2$) then
${\cal U}(\L)$ coincides with the space $H_{g, N}(1, \dots, 1)$ of  meromorphic functions of degree $N$
on Riemann surfaces of genus $g=M/2-N+1$ with $N$ simple poles and $M$ simple
critical values (see \cite{Natanzon}). The space $H_{g, N}(1, \dots, 1)$ is also called the Hurwitz space
(\cite{Natanzon}).

Following \cite{Dubrovin}, introduce the set $\hat{\cal U}(\L)$ of pairs
\begin{equation}\label{Dubrnakr}
\Big\{\L_1\in {\cal U}(\L)\ \ \Big| \ \text{a canonical basis} \ \{a_i, b_i\}_{i=1}^g\
\text{of cycles on} \ \L_1
\Big\}.
\end{equation}
The space $\hat{\cal U}(\L)$ is a covering of ${\cal U}(\L)$.

The branch points $\l_1, \dots, \l_M$ of a covering $\L_1\in {\cal U}(\L)$
can serve as local coordinates on the space ${\cal U}(\L)$ as well as on its covering
$\hat{\cal U}(\L)$.

A branched covering $\L$ is completely determined by its branch points if in addition one fixes
a representation $\sigma$ of the fundamental group $\pi_1\big({\mathbb P}^1\setminus\{\l_1, \dots, \l_M\}\big)$
in the symmetric group $S_N$. The element $\sigma_\gamma\in S_N$ corresponding to an element
$\gamma\in\pi_1\big({\mathbb P}^1\setminus\{\l_1, \dots, \l_M\}\big)$ describes the permutation of the sheets
of the covering $\L$ if the point $\l\in {\mathbb P}^1$ encircles the loop $\gamma$. One gets
a small neighborhood
of a given branched covering $\L$
moving the branch points in small neighborhoods of their initial positions
without changing  the representation $\sigma$.

\subsection{The Bergmann and Wirtinger projective connections}

 Choose on $\L$ a canonical
basis of cycles $\{a_i, b_i\}_{i=1}^g$
 and the corresponding basis of
holomorphic differentials $v_i$ normalized by the conditions
$\oint_{a_i}v_j=\delta_{ij}$. Let
\begin{equation}\label{berg}
B(P, Q)=d_Pd_Q\log E(P, Q)\;,
\end{equation}
where $E(P, Q)$ is the prime form (see \cite{Mumford} or \cite{Fay73}) ,
be the Bergmann kernel on the surface $\L$.

 The invariant Wirtinger bidifferential
$W(P, Q)$ on $\L$ is defined by the equality
\begin{equation}\label{wirtbi}
W(P, Q)=B(P, Q)+\frac{2}{4^g+2^g}\sum_{i, j=1}^g v_i(P)v_j(Q)\frac{\partial^2}{\partial z_i\partial z_j}
\log\prod_{\beta\,{\rm even}}\Theta[\beta](z\,|\,{\mathbb B})\Big|_{z=0}\;,
\end{equation}
where
${\mathbb B}=||{\mathbb B}_{ij}||_{i, j=1}^g$ is the matrix of $b$-periods of $\L$;  $\beta$ runs through the set
of all even characteristics (see \cite{Fay,Tyurin}).

In contrast to the Bergmann kernel, the invariant Wirtinger differential {\it does not} depend on the choice
of canonical basic cycles $\{a_i, b_i\}$.

The invariant Wirtinger bidifferential is not defined if the surface $\L$ has at least one
vanishing theta-constant. Thus, we introduce
the ``punctured" space ${\cal U}'(\L)\subset {\cal U}(\L)$ consisting of equivalence classes of branched coverings
with all nonvanishing theta-constants. Unless the  $g\leq2$ or $g>2$ and  $N=2$ the ``theta-divisor"
${\cal Z}={\cal U}(\L)\setminus {\cal U}'(\L)$ forms a subspace of codimension $1$ in ${\cal U}(\L)$.
If $g\leq 2$ then the set ${\cal Z}$ is empty and ${\cal U}'(\L)={\cal U}(\L)$; for hyperelliptic
($N=2$) coverings
 of genus $g>2$
a vanishing theta-constant does always exist and, therefore, for such coverings ${\cal U}'(\L)$ is empty.

The Wirtinger bidifferential has the following asymptotics near diagonal:
\begin{equation}\label{wirtas}
W(P, Q)=\left\{\frac{1}{\left(x(P)-x(Q)\right)^2}+\frac{1}{6}S_W\big(x(P)\big)+o(1)\right\}dx(P)dx(Q)
\end{equation}
as $P\rightarrow Q$, where  $x(P)$ is a local coordinate on $\L$.
The quantity $S_W$ is a projective connection on $\L$; it is called
the invariant Wirtinger projective connection.
For the Bergmann kernel we have  similar asymptotics
\begin{equation}\label{bergas}
B(P, Q)=\left\{\frac{1}{\left(x(P)-x(Q)\right)^2}+\frac{1}{6}S_B\big(x(P)\big)+o(1)\right\}dx(P)dx(Q)\;,
\end{equation}
where $S_B$ is the Bergmann projective connection. The Bergmann and the invariant Wirtinger projective
connections are related as follows:
\begin{equation}\label{Bergkon}
S_W=S_B+\frac{12}{4^g+2^g}\sum_{i, j=1}^g\left\{\frac{\partial^2}{\partial z_i\partial z_j}
\log\prod_{\beta\,{\rm even}}\Theta[\beta](z\,|\,{\mathbb B})\Big|_{z=0}\right\}v_i\,v_j\;.
\end{equation}
As well as the Wirtinger bidifferential itself, the Wirtinger
projective connection does not depend on the choice of basic cycles on
$\L$ while the Bergmann projective connection does.

We recall that any projective connection $S$ behaves as follows under the coordinate change
$x=x(z)$:
\begin{equation}\label{PSv}
S(z)=S(x)\left(\frac{dx}{dz}\right)^2+R^{x, z},
\end{equation}
where
\begin{equation}\label{Shw}
R^{x, z}\equiv\{x, z\}=\frac{x'''(z)}{x'(z)}-\frac{3}{2}\left(\frac{x''(z)}{x'(z)}\right)^2
\end{equation}
is the Schwarzian derivative.

The  following formula for the Bergmann projective connection at an
arbitrary point $P\in\L$ on the Riemann surface of genus $g\geq 1$
is a simple corollary of expression (\ref{berg}) for the Bergmann
kernel \cite{Fay73}:
\begin{equation}\label{SvFay}
S_B(x(P))=-2\frac{T}{H}+\left\{\int^P H,\; x(P)\right\},
\end{equation}
where
\ben
H=\sum\Theta^{*}_{z_i}(0)f_i;\hskip1.0cm T=\sum_{i, j, k}
\Theta^{*}_{z_i z_j
z_k}(0)f_i f_j f_k\;;
\een
$\Theta^{*}$ is the theta-function with an arbitrary non-singular odd
half-integer characteristic;  $f_i\equiv v_i(P)/dx(P) $.

\subsection{Variational formulas}

 Denote by $x_m=(\l-\l_m)^{1/r_m}$ the natural coordinate of a point $P$ in a neighborhood of the ramification
point $P_m$, where $\l=p(P)$.

Recall the Rauch   formula (see, e. g., \cite{Fay}, formula (3.21)
or the classical paper \cite{Rauchtrans}),
 which describes the variation of the
matrix ${\mathbb B}=||b_{ij}||$ of $b$-periods
 under the variation of conformal structure corresponding to a Beltrami differential $\mu\in L^\infty$:
\begin{equation}\label{varper}
\delta_{\mu}b_{ij}=\int_{\L}\mu v_i v_j\;.
\end{equation}

We shall need also the analogous formula for the variation of the Bergmann kernel
\begin{equation}\label{varberg}
\delta_{\mu}B(P, Q)=\frac{1}{2\pi i} \int_\L\mu(\,\cdot\,)B(\,\cdot\,, P)B(\,\cdot\,, Q)
\end{equation}
(see \cite{Fay}, p. 57).

Introduce the following Beltrami differential
\begin{equation}\label{sch2}
\mu_m=-\frac{1}{2\varepsilon^{r_m}}\left(\frac{|x_m|}{x_m}\right)^{r_m-2}
{\bf  1}_{\{|x_m|\leq\varepsilon\}}\frac{{d\,\bar x_m}}{d\,x_m}
\end{equation}
with sufficiently small $\varepsilon>0$
( where ${\bf  1}_{\{|x_m|\leq\varepsilon\}}$ is the function equal to $1$
inside the   disc of radius $\varepsilon$ centered at $P_m$ and vanishing outside the disc); if $r_m=2$
this Beltrami differential corresponds to the so-called
Schiffer variation).

Setting $\mu=\mu_m$ in (\ref{varper}) and using the Cauchy formula,
we get
\begin{equation}\label{beltvetvl}
\delta_{\mu_m}b_{ij}=\frac{2\pi i}{r_m\,(r_m-2)!}\left(\frac{d}{dx_m}\right)^{r_m-2}
\left\{\frac{v_i(x_m)v_j(x_m)}{(dx_m)^2}\right\}\Big|_{x_m=0}.
\end{equation}
Observe now that the r. h. s. of formula (\ref{beltvetvl}) coincides with the  known expression
for the
derivative of
the b-period with respect to the branch point $\l_m$:
\begin{equation}\label{rauchlambda}
\frac{\partial b_{ij}}{\partial \l_m}=2\pi i\,\text{res}\,\Big|_{\l=\l_m}
\sum_{k=1}^{N}\frac{1}{d\l}v_i(\l^{(k)})v_j(\l^{(k)}),
\end{equation}
where $\l^{(k)}$ denotes the point on the $k$-th sheet of the covering $\L$ which  projects
to the point $\l\in{\mathbb P}^1$.
(Only those sheets which are glued together at the point $P_m$ give a non-trivial contribution to the summation
at the r. h. s. of (\ref{rauchlambda})).
Thus, we have the following relation for variations of $b$-periods:
\begin{equation}\label{bperiod}
\partial_{\l_m} b_{ij}=\delta_{\mu_m}b_{ij}\,.
\end{equation}
This relation can be generalized for an arbitrary function of moduli.
Let $Z: T_g\rightarrow H_g$ be the standard holomorphic map from the Teichm\"uller space
$T_g$ to Siegel's generalized upper half-plane. (The $Z$ maps the conformal equivalence class of
a marked Riemann surface to the set of $b$-periods of normalized holomorphic differentials on
this surface.) It is well-known that the rank of the map $Z$ is $3g-3$ at any point
of $T_g\setminus T_g'$, where $T_g'$ is the $(2g-1)$-subvariety of $T_g$ corresponding to hyperelliptic
surfaces. Thus, one can always
choose some $3g-3\ $ $b$-periods as local coordinates
in a small neighborhood of any point of $T_g\setminus T_g'$. Using these coordinates, we get
\begin{equation}\label{osnbelt}
\frac{\delta f}{\delta\mu_m}=\sum_{i, j}\frac{\partial f}{\partial b_{ij}}\delta_{\mu_m}b_{ij}=
\frac{\partial f}{\partial \l_m},
\end{equation}
for any differentiable function $f$ on $T_g$ under the condition that the variation in the l. h. s.
of (\ref{osnbelt}) is taken at a point of $T_g\setminus T_g'$
(i. e. at a non-hyperelliptic surface).

    Formula (\ref{rauchlambda}) is well-known in the case of the simple branch point
$\l_m$ (i. e. for $r_m=2$, see, e. g., \cite{Rauch50}). Since we did not find an appropriate reference for the general case,
in what follows we briefly outline the proof:

Writing the basic differential $v_i$ in a neighborhood of the ramification point $P_m$ as
$$v_i(x_m)=\Big(C_0+C_1x_m+\dots+C_{r_m-1}x_m^{r_m-1}+O(|x_m|^{r_m})\Big)dx_m$$
and differentiating this expression with respect to $\l_m$,
we get the asymptotics
\begin{equation}\label{aslocdif}
\frac{\partial}{\partial \l_m}v_i(x_m)=
\Big\{C_0\left(1-\frac{1}{r_m}\right)\frac{1}{x_m^{r_m}}+C_1\left(1-\frac{2}{r_m}\right)\frac{1}{x_m^{r_m-1}}
+\dots +C_{r_m-2}\left(1-\frac{r_m-1}{r_m}\right)\frac{1}{x_m^2}+O(1)\Big\}dx_m.
\end{equation}
If $n\neq m$ then in a neighborhood of the ramification point $P_n$ we have the asymptotics
$$\frac{\partial}{\partial \l_m}v_i(x_n)=O(1)dx_n.$$
Therefore, the meromorphic differential $\partial_{\l_m}v_i$ has the only  pole at the point $P_m$
and its principal part at $P_m$ is given by (\ref{aslocdif}). Observe that all the $a$-periods
of $\partial_{\l_m}v_i$ are equal to zero. Thus we can reconstruct $\partial_{\l_m}v_i$ via
the first $r_m-2$  derivatives of the Bergmann kernel:
\begin{equation}\label{reconstruct}
\frac{\partial}{\partial \l_m}v_i(P)=\frac{1}{r_m(r_m-2)!}\left(\frac{d}{dx_m}\right)^{r_m-2}\Big\{
\frac{B(P, x_m)v_i(x_m)}{(dx_m)^2}\Big\}\Big|_{x_m=0}.
\end{equation}
To get (\ref{rauchlambda}) it is enough to integrate (\ref{reconstruct}) over the $b$-cycle $b_j$
(whose projection on ${\mathbb P}^1$ is independent of the branch points) and use the formula
$$\int_{b_j}B(\,\cdot\,, x_m)=2\pi i v_j(x_m).$$

 One may apply the same arguments to get the following
  formula for the derivative
of the Bergmann kernel with respect to the branch point  $\l_m$:
\begin{equation}\label{diffberg}
\frac{\partial}{\partial \l_m}B(P, Q)=-\,\text{res}\,\Big|_{\l=\l_m}\frac{1}{d\l}\sum_{k=1}^N
B(P, \l^{(k)})B(Q, \l^{(k)}).
\end{equation}
This formula also follows from (\ref{varberg}) and (\ref{osnbelt}).

We shall need also another expression for the derivative of the Bergmann kernel:
\begin{equation}\label{Rauch}
\frac{\partial}{\partial\l_m}B (P, Q)={\rm res}\big|_{\l=\l_m}\left\{\frac{1}{d\l}\sum_{j\neq k}B(P, \l^{(j)})
B(Q, \l^{(k)})\right\}\;.
\end{equation}

To prove it we note that
the sum $\sum_j B(P, \l^{(j)})$  over all the sheets of covering $\L$
 gives the Bergmann kernel on the
sphere ${\mathbb P}^1$ $$\frac{d\l d\mu(P)}{(\l-\mu(P))^2}$$
(here $\mu(P)=p(P)$), therefore, we have
\ben
\frac{(d\l)^2 d\mu(P)
d\mu(Q)}{(\l-\mu(P))^2(\l-\mu(Q))^2}=\sum_j B(P,
\l^{(j)})\sum_k B(Q, \l^{(k)})
\een
\ben
=\sum_j B(P,\l^{(j)}) B(Q, \l^{(j)})+ \sum_{j\neq k} B(P,\l^{(j)})B(Q, \l^{(k)})\;.
\een
 Now taking the residue at
$\l=\l_m$ and using (\ref{diffberg}), we get (\ref{Rauch}).

\subsection{The Bergmann and Wirtinger
projective connections
at the branch points.}
Here we prove a property of the Bergmann projective connection on a branched covering which plays
a crucial role in all our forthcoming constructions.

Introduce the following notation:
\begin{equation}\label{Hurw}
{\cal B}_m=-\frac{1}{6(r_m-2)!\,r_m}\left(\frac{d}{dx_m}\right)^{r_m-2}S_B(x_m)\big|_{x_m=0}\;,
\;\;\;m=1, 2, \dots, M,
\end{equation}
where $S_B(x_m)$ is the Bergmann projective connection corresponding to the local parameter
$x_m=(\l-\l_m)^{1/r_m}$ near the ramification point $P_m$.
(The factor
$-\frac{1}{6}$ in (\ref{Hurw}) seems to be of no importance, its appearance will be explained later on.)

If we deform  covering (\ref{nakr})  moving the branch points in small neighborhoods of their
initial positions and preserving the permutations corresponding to the branch points
then the
quantity ${\cal B}_m$ becomes
a function of $(\l_1, \dots, \l_M)$.

\begin{theorem}\label{proizv}
For any $m, n=1, \dots, M$ the following equations hold
\begin{equation}\label{cross}
\frac{\partial{\cal B}_m}{\partial \l_n}=\frac{\partial{\cal B}_n}{\partial \l_m}.
\end{equation}
\end{theorem}

{\bf Proof.} We start with the following lemma.
\begin{lemma}
The function ${\cal B}_m$ can be expressed via the Bergmann kernel as
\begin{equation}\label{Bersv}
{\cal B}_m=2\, {\rm res} \big|_{\l=\l_m}\left\{\frac{1}{d\l}\sum_{k, j=1; j\neq k}^N
B(\l^{(j)}, \l^{(k)})\right\},
\end{equation}
where $\l^{(j)}$ is the point of the $j$-th sheet of covering (\ref{nakr}) such that $p(\l^{(j)})=\l$.
\end{lemma}

Let $H(\,\cdot\, , \,\cdot\,)$ be the nonsingular part of the Bergmann kernel, i. e.
$$B(P, Q)=\left(\frac{1}{(x(P)-x(Q))^2}+H(x(P), x(Q))\right)dx(P)\,dx(Q),$$
as $P\,\to\,Q$.

To prove the lemma we observe that only those sheets which are glued together at the point $P_m$
give a non-trivial contribution to the summation in (\ref{Bersv}). Now we may
rewrite the right hand side  of (\ref{Bersv}) as
\ben
\frac{1}{3}\,{\rm res}\big|_{\l=\l_m} \sum_{j,k=1,\, j\neq k}^{r_m}
H(\g^jx_m,\g^kx_m)\g^{j+k} \left(\f{dx_m}{d\l}\right)^2\,d\l\;,
\een
where $\g=e^{{2\pi i}/{r_m}}$ is the root of unity. In terms of
coefficients of the Taylor series of $H(x_m, y_m)$ at the point
$P_m$:
\ben
H(x_m, y_m)=\sum_{s=0}^{\infty}\sum_{p=0}^s
\f{H^{(p,s-p)}(0,0)}{p! (s-p)!}x_m^p y_m^{s-p}
\een
 this expression
looks as follows:
 \ben \f{1}{3r_m^2}\sum_{p=0}^{r_m-2}
\f{H^{(p,r_m-2-p)}(0,0)}{p! (r_m-2-p)!}
\sum_{j,k=1\,,\; j<k}^{r_m} \g^{(p+1)k + (r_m-p-1)j}\; .
\een
Summing up the
geometrical progression, we get (\ref{Bersv}).

Using
(\ref{Bersv}) and (\ref{Rauch}) we conclude that
\ben
\frac{\partial{\cal B}_m}{\partial\l_n}=2\left\{
\frac{\partial}{\partial\l_n}\text{res}\big|_{\l_m}
\frac{1}{d\l}\sum_{j\neq k}B(\l^{(j)}, \l^{(k)})\right\}=
\een
\ben
=2\,
\text{res}\big|_{\l=\l_m}\text{res}\big|_{\mu=\l_n}\left\{\frac{1}{d\l}\frac{1}{d\mu}
\sum_{j\neq  k}\sum_{j'\neq k'}B(\mu^{(j')},
\l^{(j)})B(\mu^{(k')}, \l^{(k)})\right\}\;.
\een
 To finish the
proof we note that the last expression is symmetric with respect
to $m$ and $n$.
$\square$

The analogous statement is also true for the derivatives of the Wirtinger projective connection.
Namely, set
\begin{equation}\label{SvWirt}
{\cal A}_m=-\frac{1}{6(r_m-2)!\,r_m}\left(\frac{d}{dx_m}\right)^{r_m-2}S_W(x_m)\big|_{x_m=0}\;,
\;\;\;m=1, 2, \dots, M,
\end{equation}
where $S_W(x_m)$ is the Wirtinger projective connection corresponding to the local parameter
$x_m$ near the ramification point $P_m$. The following statement is an easy corollary
of Theorem \ref{proizv}.

\begin{theorem}\label{proizv2}
For any $m, n=1, \dots, M$ the following equations hold
\begin{equation}\label{cross2}
\frac{\partial{\cal A}_m}{\partial \l_n}=\frac{\partial{\cal A}_n}{\partial \l_m}.
\end{equation}
\end{theorem}

{\bf Proof.} A simple calculation shows that the one-form
 $${\cal V}=\sum_{m=1}^M({\cal A}_m-{\cal B}_m)d\l_m\;$$
 is a total differential:
\begin{equation}\label{dobavka}
{\cal V}=-\frac{4}{4^g+2^g}d\log\prod_{\beta\,{\rm even}}\Theta[\beta](0\,|\,{\mathbb B})\;.
\end{equation}
To prove (\ref{dobavka}) it is sufficient to use
the heat equation for theta-function
\begin{equation}
\f{\p \Theta[\beta](z\,|\,{\mathbb B})}{\p b_{jk}}=
\f{1}{4\pi i}\f{\p^2 \Theta[\beta](z\,|\,{\mathbb B})}{\p z_j \p z_k}\;,
\end{equation}
the  formula (\ref{beltvetvl}) for the derivative of the $b$-period with respect to the branch point
and the link (\ref{Bergkon}) between the Wirtinger and Bergmann projective connections. $\square$

\subsection{The Wirtinger and Bergmann tau-functions of branched coverings}
\subsubsection{The Wirtinger tau-function}
We recall that ${\cal U}'(\L)$ denotes the set of branched coverings
from the connected component ${\cal U}(\L)\ni \L$ of the Hurwitz space $H(N, M, {\mathbb P}^1)$
for which none of the
theta-constants vanishes.
Introduce the connection
\begin{equation}\label{connect1}
d_W=d-\sum_{m=1}^M{\cal A}_md\l_m\;,
\end{equation}
acting in the trivial bundle ${\cal U}'(\L)\times{\mathbb
C}$, where $d$
is the external differentiation (having both ``holomorphic" and ``antiholomorphic" components);
the connection coefficients ${\cal A}_m$ are
defined by (\ref{SvWirt}).

\begin{remark}\rm{
If we choose another global holomorphic coordinate $\tilde\lambda$ on ${\mathbb P}$,
$
\l=(a\tilde\lambda+b)/(c\tilde\lambda+d)\;,
$
where $ad-bc=1$, then the connection $d_W$ turns into a gauge equivalent connection.
Consider, for example, the case of branched coverings with simple branch points (all the $r_m$ are equal to $2$).
Let $\tilde\lambda_m$ be the new coordinates of the branch points,
\begin{equation}\label{Mobius}
\l_m=\frac{a\tilde\lambda_m+b}{c\tilde\lambda_m+d}\;;
\end{equation}
then the gauge transformation of connection $d_W$ in local coordinates looks as follows
\begin{equation}\la{gauge}
d_W \mapsto G^{-1}d_W G\;,
\end{equation}
where
\begin{equation}\la{gauge1}
G=\prod_{m=1}^M(c\tilde\lambda_m+d)^{-1/4}\;.
\end{equation}}
\end{remark}

Theorem \ref{proizv2} implies the following statement.

\begin{theorem}\label{plosk}
The connection  $d_W$, defined in the trivial line bundle over
${\cal U}'(\L)$ in terms of the Wirtinger projective connection by
formulas (\ref{connect1}), (\ref{SvWirt}), is flat.
\end{theorem}

The flat connection $d_W$ determines a character of the fundamental group of ${\cal U}'(\L)$ i.e. the
representation
\begin{equation}\label{repr}
\rho:\pi_1\big({\cal U}'(\L)\big)\rightarrow {\mathbb C}^*\;.
\end{equation}

Denote by  ${\cal E}$  the universal covering of ${\cal U}'(\L)$; then
the group $\pi_1\big({\cal U}'(\L)\big)$ acts on the direct product
${\cal E}\times {\mathbb C}$ as follows:
$$g(e, z)=(ge, \rho(g)z)\;,$$
where $e\in {\cal E}$, $z\in {\mathbb C}$, $g\in \pi_1\big({\cal U}'(\L)\big)$.
The factor manifold $\;{\cal E}\times{\mathbb C}/\pi_1\big({\cal U}'(\L)\big)$
has the  structure
of a holomorphic line bundle over ${\cal U}'(\L)$; we denote this bundle by ${\cal T}_W$.
\begin{definition}\rm
The flat holomorphic line bundle ${\cal T}_W$ equipped with the flat connection $d_W$ is called
the Wirtinger line bundle over the punctured Hurwitz space ${\cal U}'(\L)$.
The (unique up to a multiplicative constant) horizontal holomorphic
section of the bundle  ${\cal T}_W$ is called
the Wirtinger $\tau$-function of the covering $\L$ and denoted by $\tau_W$.
\end{definition}

Taking into account the form (\ref{gauge}), (\ref{gauge1}) of the
gauge transformation of connection $d_W$ under conformal
transformations on the base $\l$-plane, we see that the Wirtinger
tau-function $\tau_W$ of a branched covering with simple branch points
transforms as follows under conformal transformation (\ref{Mobius}):
\begin{equation}\label{prostgauge}
\tau_W\mapsto \prod_{m=1}^M(c\tilde\lambda_m+d)^{-1/4} \tau_W\;.
\end{equation}
One can easily derive the analogous formula in the general case of an arbitrary covering.

We notice that
\begin{itemize}
\item In genera $0$, $1$ and $2$ the ``theta-divisor" ${\cal Z}={\cal U}(\L)\setminus{\cal U}'(\L)$
is empty. Therefore, in this case the bundle ${\cal T}_W$ is a  bundle over the whole connected component
${\cal U}(\L)$ of the  Hurwitz space $H(N, M, {\mathbb P}^1)$.
\item
Hyperelliptic coverings $(N=2)$  fall within this framework only in genera $g=0,1,2$
since for genus $g>2$ one of the theta-constants always
vanishes for hyperelliptic curves \cite{Mumford}.
\item In the case of simple branch points the space ${\cal U}(\L)$ is nothing but the Hurwitz space
$H_{g, N}(1, \dots, 1)$ from (\cite{Dubrovin}, \cite{Natanzon}).
\end{itemize}
\subsubsection{The Bergmann tau-function}

Consider now the covering $\hat{\cal U}(\L)$  (the set of pairs (\ref{Dubrnakr}))
of the space ${\cal U}(\L)$.
Repeating the construction of the previous subsection for the flat connection
\begin{equation}\label{connect2}
d_B=d-\sum_{m=1}^M{\cal B}_md\l_m\;,
\end{equation}
in the trivial line bundle $\hat{\cal U}(\L)\times{\mathbb C}$,
we get a flat holomorphic line bundle ${\cal T}_B$ over $\hat{\cal U}(\L)$.

(Here the coefficients ${\cal B}_m$ are defined by formula (\ref{Hurw}), the flatness of connection
(\ref{connect2}) follows from Theorem \ref{proizv}.)

\begin{definition}\rm
The flat holomorphic line bundle ${\cal T}_B$ equipped with the flat connection $d_B$ is called
the Bergmann line bundle over the covering  $\hat{\cal U}(\L)$
of the connected component ${\cal U}(\L)$ of the Hurwitz space $H(N, M, {\mathbb P}^1)$.
The (unique up to a multiplicative constant) horizontal holomorphic
section of the bundle  ${\cal T}_B$ is called
the Bergmann $\tau$-function of the covering $\L$ and denoted by $\tau_B$.
\end{definition}

According to the link (\ref{Bergkon}) between Wirtinger and Bergmann
projective connections, the corresponding tau-functions are related as follows:
\begin{equation}\la{svBergWirt}
\tau_W=\tau_B\left\{\prod_{\beta\,{\rm even}}\Theta[\beta](0\,|\,{\mathbb B})\right\}^{-\frac{1}
{4^{g-1}+2^{g-2}}}\;.
\end{equation}
In contrast to the Wirtinger tau-function, the Bergmann tau-function does depend upon
the choice of canonical basis of cycles on $\L$.

Consider the case of hyperelliptic (N=2) coverings.
As a by-product of  computation of
isomonodromic tau-functions for  Riemann-Hilbert problems with quasi-permutation monodromies (see \cite{KitKor}),
it was found the following expression for the
Bergmann tau-function $\tau_B$ on the spaces $\hat{H}_{g,2}(1, 1)$:
\begin{equation}
\tau_B= \det {\cal A} \prod_{m, n =1; \ m<n}^{2g+2}(\l_m-\l_n)^{1/4}\;,
\la{Btauhe}
\end{equation}
where  ${\cal A}$ is the matrix of $a$-periods of non-normalized holomorphic differentials on $\L$:
$ {\cal A}_{\a\b} =\oint_{a_\a}\f{\l^{\b-1}d\l}{\nu}$, with $\nu^2=\prod_{m=1}^{2g+2} (\l-\l_m)$.

Expression (\ref{Btauhe}) coincides with the empirical formula for the
determinant of  $\bar{\partial}$-operator, acting in the trivial line
bundle over $\L$, derived in \cite{Knizhnik}.
Due to the term $\det {\cal A}$,  the expression (\ref{Btauhe}) is explicitly dependent on the choice of canonical basis of cycles on
$\L$.

On the other hand, the Wirtinger tau-function, which is independent  of the choice of canonical basis of cycles,
is defined on hyperelliptic curves only if $g\leq 2$. Consider the case $g=2$ (postponing the
cases $g=0, 1$ to the next section).

Recall the classical Thomae formulas, which express the  theta-constants of hyperelliptic curves in terms of branch points.
Namely, consider an arbitrary partition of the set of branch points $\{\l_1,\dots,\l_{2g+2}\}$ into two subsets: $T$ and
$\overline{T}$, where the  subset $T$ (and also $\overline{T}$) contains $g+1$ branch points. To each such partition we can associate
an even vector of  half-integer characteristics $[\eta'_T,\eta''_T]$ such that
\begin{equation}
{\mathbb B}\eta'_T+\eta''_T = \sum_{\l_m\in T} U(\l_m) - K \;,
\end{equation}
where $U(P)$ is the Abel map, $K$ is the vector of Riemann constants.
The number of even characteristics obtained in this way is given by
$\f{1}{2}C_{2g+2}^{g+1}$.
If we denote the  theta-function with characteristics
 $[\eta'_T,\eta''_T]$ by $\theta[\beta_T]$, the Thomae formula (see \cite{Mumford}) states that related theta-constant
can be computed as follows:
\begin{equation}
\Theta^4[\beta_T](0) = \pm(\det{\cal A})^2 \prod_{\l_m,\l_n\in T} (\l_m-\l_n)\prod_{\l_m,\l_n\in \overline{T}} (\l_m-\l_n)\;.
\la{Thomae}
\end{equation}

In genus $2$ we have $\f{1}{2}(4^2 + 2^2)=10$ even characteristics in total; this number coincides with the number
$\f{1}{2}C_6^3$ of non-vanishing  even characteristics for which the Thomae formulas take place.
Substitution of Thomae formulas (\ref{Thomae}) and expression (\ref{Btauhe}) for $\tau_B$ into (\ref{svBergWirt}) gives the
following formula for the Wirtinger tau-function of a hyperelliptic covering of genus $2$:
\begin{equation}
\tau_W=\prod_{m,n =1,  \ m<n}^6(\l_m-\l_n)^{\frac{1}{20}}\;.
\la{tau22}
\end{equation}
The independence of the Wirtinger tau-function of the choice of canonical basis of cycles on $\L$
is manifest here.

\begin{remark}{\rm
For higher  genus ($g>2$) two-fold coverings our definition of Wirtinger tau-function
does not work, since
some of theta-constants always vanish.
However, we can slightly  modify formula (\ref{svBergWirt}), averaging only over the set of non-singular
even characteristics. This leads to the following definition
\begin{equation}
\tau_W^*= \tau_B\left\{\prod_{T}\Theta[\beta_T](0|{\mathbb B})\right\}^{-{4}/C_{2g+2}^{g+1}}\;.
\la{wirthe}
\end{equation}
Since the set of all  characteristics $\beta_T$ is invariant with respect to any change of canonical basis of cycles,
function $\tau_W^{*}$ does not depend on the choice of this basis.
Substitution of expression (\ref{Btauhe}) and Thomae formulas
(\ref{Thomae}) into (\ref{wirthe}) leads to the following result:
\begin{equation}
\tau_W^{*}=\prod_{m,n=1\; m\neq n}^{2g+2}(\l_m-\l_n)^{1/4(2g+1)}\;.
\la{tauheg}
\end{equation}}
\end{remark}

The main goal of the present paper is the calculation of the Wirtinger and Bergmann tau-functions
of {\it an arbitrary covering} $\L$.
In section \ref{REC} we explicitly calculate them
for coverings of genera $0$ and $1$. For arbitrary coverings of higher genus we are able to calculate only
the modulus square of the tau-function (see section  \ref{SHG}).

\section{Rational and elliptic cases}\label{REC}

If $g=0$ the branched covering $\L$
can be biholomorphically mapped to the Riemann sphere ${\mathbb P}^1$.
Let $z$ be the natural coordinate on
${\mathbb P}^1\setminus\infty$.
 The projective connection $S_B(x_m)$
 reduces to the Schwarzian derivative
\ben
S_B(x_m)=R^{z, x_m}=\{z(x_m), x_m\}\;.
\een
Therefore
\begin{equation}\label{sph1}
{\cal B}_m=\frac{-1}{6r_m\,(r_m-2)!}\left(\frac{d}{dx_m}\right)^{r_m-2}R^{z, x_m}|_{x_m=0}\;.
\end{equation}

If $g=1$ the branched covering $\L$ can be biholomorphically mapped to the torus with periods $1$ and $\mu$;
in genus $1$ there is only one theta-function with odd characteristic
which is the odd Jacobi theta-function $\theta_1(z|\mu)=\theta\left[^{1/2}_{1/2}\right](z\,|\,\mu)$.
 Using
(\ref{SvFay}) and the heat equation $\partial_z^2{\theta_1}=4\pi i\partial_\mu\theta_1$,
 we get
\ben
S_B(x_m)=-8\pi i \frac{\partial\log
{\theta_1}'}{\partial \mu}v^2(x_m)+R^{z, x_m}\;,
\een
 where $\theta_1'\equiv\partial\theta_1/\partial z|_{z=0}$,
$v=v(x_m)dx_m$ and $z=\int^Pv$. Now the variational formula (\ref{beltvetvl})
implies that
\begin{equation}\label{Hurtor}
{\cal B}_m=\frac{2}{3}\frac{\partial\ln{\theta_1}'}{\partial\lambda_m}
-\frac{1}{6r_m\,(r_m-2)!}\left(\frac
{d}{dx_m}\right)^{r_m-2}R^{z, x_m}|_{x_m=0}\;.
\end{equation}

Our way of calculating of the tau-functions $\tau_W$ and $\tau_B$ is rather indirect. Namely,
 we shall first compute the module of the tau-function.
Since the first term in (\ref{Hurtor}) can be immediately
integrated, in both  cases $g=0$ and $g=1$ one needs to find
a real-valued potential ${\mathbb S}(\lambda_1, \dots, \lambda_n)$ satisfying
\begin{equation}\label{osnovn}
\frac{\partial {\mathbb S}}{\partial\l_m}=
\frac{1}{(r_m-2)!\,r_m}\left(\frac{d}{dx_m}\right)^{r_m-2}
R^{z, x_m}|_{x_m=0}\;,
\end{equation}
where $z$ is the natural coordinate on the universal covering of $\L$
(i.e. on the complex plane for $g=1$ and the Riemann sphere for $g=0$).

The solution of equations (\ref{osnovn}) is given by  Theorem 4 below.
The function ${\mathbb S}$ turns out to coinside
with the properly regularized Dirichlet integral
\begin{equation}\label{Liact}
\frac{1}{2\pi}\int_\L|\phi_\l|^2,
\end{equation}
where  $e^{\phi}|d\l|^2$ is the flat
metric on $\L$ obtained by projecting  the standard metric
$|dz|^2$
from  the universal covering.  (In case $g=0$, when the
universal covering is the Riemann sphere,  the metric $|dz|^2$ is
singular.)

The Dirichlet integral
(\ref{Liact}) can be explicitly represented as  the  modulus square
of holomorphic function of variables $\l_1, \dots, \l_M$. The
procedure of holomorphic factorization gives us the value of the
tau-function itself.

The next two subsections are devoted to the calculation of the function
${\mathbb S}$.

\subsection{The flat metric on Riemann surfaces of genus $0$ and $1$}

{\bf The asymptotics of the flat metric near the branch points.}
Compact Riemann surfaces $\L$ of genus $1$ and $0$ have the universal coverings
$\tilde\L={\mathbb C}$ and $\tilde\L={\mathbb P}^1$ respectively.
Projecting from the universal covering onto $\L$ the metric $|dz|^2$,
 we obtain the metric
of the Gaussian curvature $0$ on $\L$. (In case $g=0$ the obtained metric has singularity at the image of
the infinity of ${\mathbb P}^1$).
Let $J: \tilde\L\rightarrow \L$ be the uniformization map; denote its
inverse by  $U=J^{-1}$. Denote by  $x$ a
local parameter on $\L$.
The projection of the metric $|dz|^2$ on $\L$ looks as follows:
\begin{equation}\label{1.2}
e^{\phi(x, \bar{x})}|dx|^2=|U_{x}(x)|^2 |dx|^2;
\end{equation}
where the function $\phi$ satisfies the Laplace equation
\begin{equation}\label{Li}
\phi_{x\bar{x}}=0.
\end{equation}

In the case $g=1$ the map $P\mapsto U(P)$ may be defined by
$$U(P)=\int^P v$$
with any holomorphic differential $v$ on $\L$ (not necessarily  normalized).

In the case $g=0$ we choose one sheet of the covering $\L$ (we shall call this sheet the first one)
and require that $U(\infty^{(1)})=\infty$,
where $\infty^{(1)}$ is the infinity of the first sheet.

Choose any
sheet of the covering $\L$ (this will be a copy of the Riemann sphere ${\mathbb P}^1$
with appropriate cuts between the branch points; we recall that it is assumed
that the infinities of all the sheets are not the ramification points)
and cut out  small neighborhoods of all
the branch points and a neighborhood of the infinity. In the remaining domain we can use $\l$
as global coordinate.
Let $\phi^{ext}(\lambda, \bar\lambda)$
be the function from  (\ref{1.2})  corresponding to the coordinate $x=\lambda$ and $\phi^{int}(\lp, {\bar x}_m)$
be the function from  (\ref{1.2})  corresponding to the coordinate $x=x_m$.
\begin{lemma}
The derivative of the function $\phi^{ext}$ has the following asymptotics near the branch points and the
infinities of the sheets:
\begin{enumerate}
\item $|\phi_\lambda^{ext}(\lambda, \lambda)|^2=(\frac{1}{r_m}-1)^2|\lambda-\l_m|^{-2}+O(|\lambda-\l_m|^{-2+1/r_m
})$
as $\lambda\rightarrow \l_m$,
\item $|\phi_\lambda^{ext}(\lambda, \lambda)|^2=4|\lambda|^{-2}+O(|\lambda|^{-3})$ as $\lambda\rightarrow \infty$.
\item In the case $g=0$ on the first sheet the last asymptotics is replaced by
$$|\phi_\lambda^{ext}(\lambda, \lambda)|^2=O(|\l|^{-6})$$
as $\l\to\infty$.
\end{enumerate}
\end{lemma}
{\bf Proof.} In a small punctured neighborhood of $P_m$ on the chosen sheet we have
\begin{equation}\label{taulambda}
e^{\phi^{int}(\lp, {\bar x}_m)}|d\lp|^2=e^{\phi^{ext}(\lambda, \bar\lambda)}|d\lambda|^2.
\end{equation}
This gives the equality
$$e^{\phi^{ext}(\lambda, \bar\lambda)}=\frac{1}{r_m^2}e^{\phi^{int}(\lp, {\bar x}_m)}|\lambda-\l_m|^{2/r_m-2}$$
which implies the first asymptotics.

In a neighborhood of the infinity of the chosen sheet we may introduce the coordinate $\zeta=1/\lambda$.
Denote by $\phi^\infty(\zeta, \bar\zeta)$ the function $\phi$ from (\ref{1.2}) corresponding to the coordinate
$w=\zeta$.
 Now the second asymptotics follows from the equality
\begin{equation}\label{infty1}
e^{\phi^{ext}(\lambda, \bar\lambda)}=e^{\phi^\infty(\zeta, \bar\zeta)}|\lambda|^{-4}.
\end{equation}

In the case $g=0$ near the infinity of the first sheet we have
$$U(\l)=c_{1}\l+c_0+c_{-1}\frac{1}{\l}+\dots$$
with $c_{1}\neq0$. So at the infinity of the first sheet there is the asymptotics
$$\phi^{ext}_\l(\l, \bar\l)=\frac{U_{\l\l}}{U_\l}=O(|\l|^{-3}).$$
$\square$

{\bf The Schwarzian connection in terms of the flat metric.}
Let $x$ be some local coordinate on $\L$. Set $z=U(x)$; here $z$ is a point of the universal covering (${\mathbb C}$ or ${\mathbb P}^1$).
 The system
of Schwarzian derivatives $R^{z,x}$ (each derivative corresponds to its own local chart)
forms a projective connection on the surface $\L$. In accordance with \cite{HawSch},
we  call it the Schwarzian connection.
\begin{lemma}
\begin{enumerate}
\item
The Schwarzian connection can be expressed as follows in terms of the function $\phi$ from
 (\ref{1.2}):
\begin{equation}\label{Agregat}
R^{z, x}=\phi_{xx}-\frac{1}{2}\phi_x^2\ \,.
\end{equation}
\item
In a neighborhood of a branch point $P_m$  there is the following relation between
the values of Schwarzian connection computed with respect to coordinates $\l$ and $x_m$:
\begin{equation}\label{Agrpoint}
R^{z, \lambda}=\frac{1}{r_m^2}(\lambda-\l_m)^{2/r_m-2}R^{z, \lp}+\left(\frac{1}{2}-\frac{1}{2r_m^2}\right)(\lambda-
\l_m)^{-2}.
\end{equation}
\item
Let $\zeta$ be the coordinate in a neighborhood of the infinity of any sheet of covering (\ref{nakr})
(except the first one in the case $g=0$), $\zeta=1/\lambda$.
Then
 \begin{equation}\label{Agrinf}
R^{z, \lambda}=\frac{R^{z, \zeta}}{\l^4}=O(|\lambda|^{-4}).
\end{equation}
\end{enumerate}
\end{lemma}

{\bf Proof.} The second and the third statements are just the rule of transformation of the Schwarzian derivative
under the coordinate change. The formula (\ref{Agregat}) is well-known and
can be verified by a straightforward calculation.

$\square$

{\bf The derivative of the metric with respect to a branch point.}
In this item we set $\phi(\lambda, \bar\lambda)=\phi^{ext}(\lambda, \bar\lambda)$. The following lemma describes
the dependence of the function $\phi$ on positions of the branch points of the covering $\L$.
\begin{lemma}Let $g=0, 1$. The derivative of the function $\phi$ with respect to $\l$ is related to
its derivative with respect to a branch point $\l_m$ as follows:
\begin{equation}\label{QuaziAlfors}
\frac{\partial\phi}{\partial\lambda_m}+F_m\frac{\partial \phi}{\partial \lambda}+
\frac{\partial F_m}{\partial\lambda}=0\;,
\end{equation}
where
\begin{equation}\label{koeff}
F_m=-\frac{U_{\lambda_m}}{U_\lambda}\;.
\end{equation}
\end{lemma}

{\bf Proof.}  We have
$
\phi=\log U_\lambda +\log\overline{U_\lambda}
$;
$\phi_\lambda=\frac{U_{\lambda\lambda}}{U_\lambda}$,
$\phi_{\lambda_m}=\frac{U_{\l\l_m}}{U_\l}$ and
$$\frac{U_{\l\l_m}}{U_\l}=\frac{U_{\l_m}}{U_\l}\frac{U_{\l\l}}{U_\l}+\left(\frac{U_{\l_m}}{U_\l}\right)_\l.$$
(We used the fact that the map $U$ depends on the branch points
holomorphically.)

$\square$

\begin{lemma}
Let $g=0$ or $g=1$ and let $J$ be the uniformization map $J:{\mathbb
C}P^1  \rightarrow \L$
or $J: {\mathbb C} \rightarrow \L$ respectively.
Denote the composition
$p\circ J$ by $R$. Then
\begin{enumerate}
\item
The following relation holds:
\begin{equation}\label{w}
F_m=\frac{\partial R}{\partial \lambda_m}.
\end{equation}

\item
In a neighborhood of the branch point $\lambda_l$ the following
 asymptotics holds:
\begin{equation}\label{c}
F_m=\delta_{lm}+o(1),
\end{equation}
where $\delta_{lm}$ is the Kronecker symbol.
\item
At the infinity of each sheet (except the first sheet for  $g=0$)
the following asymptotics holds:
\begin{equation}\label{d}
F_m(\lambda)=O(|\lambda|^2).
\end{equation}
\end{enumerate}
\end{lemma}

{\bf Proof.} Writing the dependence on the branch points explicitly we have
\begin{equation}\label{tozd}
U(\lambda_1, \dots, \lambda_M; R(\lambda_1, \dots, \lambda_M; z))=z
\end{equation}
for any $z$ from the universal covering (${\mathbb P}^1$ for $g=0$ or ${\mathbb C}$ for $g=1$). Differentiating (\ref{tozd}) with respect to $\lambda_m$
we get (\ref{w}).

Let $z_0=z_0(\lambda_1, \dots, \lambda_M)$ be a point from
  the universal covering such that $J(z_0)=
P_m$. The map $R$ is holomorphic
and in a neighborhood of $z_0$ there is the representation
\begin{equation}\label{vetvl}
R(z)=\lambda_m+(z-z_0)^{r_m}f(z, \lambda_1, \dots, \lambda_M)
\end{equation}
with some holomorphic function $f(\cdot, \lambda_1, \dots, \lambda_M)$. This together with the first statement of the lemma
give (\ref{c}).

 Let now $z_\infty=z_\infty(\lambda_1, \dots, \lambda_M)$ be a point from the universal covering
 such that $J(z_\infty)=\infty$, where $\infty$ is the infinity of the chosen sheet. Then in a neighborhood of $z_\infty$ we have
$$\lambda=R(z)=g(z; \lambda_1,\dots, \lambda_M)(z-z_\infty)^{-1}$$
with holomorphic $g(\cdot, \lambda_1, \dots, \lambda_M)$. Using the first statement of the lemma,  we get (\ref{d}).

$\square$
\vskip0.5cm
\begin{corollary}\label{kratnost} Keep $m$ fixed and define $\Phi_n(x_n)\equiv F_m(\lambda_n+x_n^{r_n})$.
Then
\ben
\Phi_n(0)=\delta_{nm};\ \ \ \ \ \
\left(\frac{d}{dx_n}\right)^k\Phi_n(0)=0,\ \ \  k=1, \dots, r_n-2\;.
\een
\end{corollary}
This immediately follows from formulas (\ref{w}) and (\ref{vetvl}).

Formulas (\ref{QuaziAlfors}) and (\ref{c}) are analogous to the Ahlfors lemma as it was formulated in
\cite{ZT2}. However, they are more elementary, since their proof does not use Teichm\"uller's
theory.

\subsection{The regularized Dirichlet integral}
We recall that the covering $\L$ has $N$ sheets and $N=\sum_{m=1}^M(r_m-1)/2-g+1$ due to
the Riemann-Hurwitz formula.
To the $k$-th sheet $\L_k$ of the covering $\L$  there corresponds
 the function $\phi_k^{ext}: \L_k\to {\mathbb R}$ which
is smooth in any domain $\Omega^k_r$
of the form $\Omega^k_\rho=\{\lambda\in \L_k: \forall m \;
|\lambda-\lambda_m|>\rho\;\; \& \;\;  |\lambda|<1/\rho\}$, where $\rho>0$.
 Here $\lambda_m$ are all
the branch points which belong to  the $k$th sheet $\L_k$ of $\L$.
In the case of genus zero the above definition of the domain $\Omega^k_\rho$ is valid for $k=2, \dots, N$.
The domain $\Omega^1_\rho$ in this case should be defined separately:
\ben
\Omega^1_\rho=\{\lambda\in \L_1\setminus\infty^{1}: \forall m \;
|\lambda-\lambda_m|>\rho\}\;.
\een
(Here, again, $\l_m$ are all the branch points from the first sheet.)
We recall that in the case $g=0$ we have singled out one sheet of the covering (the first sheet
in our enumeration).
The function $\phi_k^{ext}$ has finite limits at the cuts (except the endpoints which are the ramification
 points);
at the ramification points and at infinity it possesses the asymptotics listed in Lemma 3.

Let us introduce the regularized Dirichlet integral
$$\frac{1}{2\pi}\int_\L|\phi_\l|^2\, dS.$$
Namely, set
\begin{equation}\label{action}
Q^\rho=\sum_{k=1}^N\int_{\Omega_\rho^k}|\partial_\lambda\phi^{ext}_k|^2
\,dS\;,
\end{equation}
where $dS$ is the area element on ${\mathbb C}^1$:
$
dS=\frac{|d\l\wedge d\bar\l|}{2}$.

According to Lemma 3 there exist the finite limits
\begin{equation}\label{reg}
{\mathbb S}_{ell}(\l_1, \dots, \l_M)=\frac{1}{2\pi}\lim_{\rho\to 0}
\left(Q^\rho+\Big\{4N+\sum_{m=1}^M\frac{(r_m-1)^2}{r_m}\Big\}2\pi\log \rho\right)+\sum_{m=1}^M(1-r_m)\ln r_m
\end{equation}
in the case $g=1$
and
\begin{equation}\label{sph}
{\mathbb S}_{rat}(\l_1, \dots, \l_M)=\frac{1}{2\pi}\lim_{\rho\to 0}\left(
Q_\rho+
\Big\{4(N-1)+\sum_{m=1}^M\frac{(r_m-1)^2}{r_m}\Big\}2\pi\log \rho\right)+\sum_{m=1}^M(1-r_m)\ln r_m
\end{equation}
in the case $g=0$;
the last constant term $\sum_{m=1}^M(1-r_m)\ln r_m$ we include for convenience.

\begin{theorem}\label{torsph}
Let ${\mathbb S}={\mathbb S}_{rat}$ for $g=0$, ${\mathbb S}={\mathbb S}_{ell}$ for
$g=1$. Then for any $m=1, \dots, M$
\begin{equation}\label{torus}
\frac{\partial{\mathbb S}(\l_1, \dots, \l_M)}{\partial \l_m}=
\frac{1}{(r_m-2)!\,r_m}\left(\frac{d}{dx_m}\right)^{r_m-2}
R^{z, x_m}|_{x_m=0}\;,
\end{equation}
where $z$ is the natural coordinate on the universal covering of $\L$
(${\mathbb P}^1$ for $g=0$ and ${\mathbb C}$ for $g=1$).
\end{theorem}

{\bf Proof.}
We shall restrict ourselves to the case $g=1$. The proofs for $g=0$ and $g=1$ differ only in details
concerning the infinity of the first sheet.

Let $Q^\rho$ be defined by formula (\ref{action}).
 We have
\begin{equation}\label{osn}
\frac{\partial}{\partial \lambda_m}Q^\rho=\frac{i}{2}\sum_{l=1}^{r_m}\oint_{|\lambda^{(l)}-\lambda^{(l)}_m|=\rho}
|\partial_\lambda\phi|^2d\bar\lambda+\sum_{k=1}^N\int\int_{\Omega_\rho^{(k)}}\frac{\partial}
{\partial\lambda_m}|\partial_\lambda \phi|^2dS\;.
\end{equation}
Here the first sum corresponds to those sheets of the covering (\ref{nakr}) which are glued together at the point
$P_m$; the upper index $(l)$  signifies that the integration
is over a contour lying on the $l$-th sheet.
\begin{lemma}\label{svyazn}
There is an equality
\begin{equation}\label{lem}
\frac{2}{(r_m-2)!\,r_m}\left(\frac{d}{d x_m}\right)^{r_m-2}R^{z, x_m}\big|_{x_m=0}=
-\sum_{n=1}^M \left(1-\frac{1}{r_n^2}\right)\frac{1}{(r_n-1)!}\left(\frac{d}{dx_n}\right)^{r_n}
F_m(\lambda_n+{x_n}^{r_n})\big|_{x_n=0}.
\end{equation}
Here $x_n$, $x_m$ are the local parameters near $P_n$ and $P_m$.
The summation at the right is over all the branch points of the covering $\L$.
\end{lemma}

{\bf Proof.}  Using (\ref{Agregat}) and the holomorphy of $R^{z, \lambda}$ with respect to $\lambda$, we have
\begin{equation}\label{x.1}
0=\sum_{k=1}^N\oint_{\partial \Omega_\rho^k} F_m(2\phi_{\lambda\lambda}-\phi_\lambda^2)\,d\lambda=
2\sum_{k=1}^N\oint_{|\lambda|=1/\rho}F_m R^{z, \lambda}\,d\lambda+\sum_{k=1}^N\sum_{\lambda_n \in \L_k}
\oint_{|\lambda-\lambda_n|=\rho}F_m(2\phi_{\lambda\lambda}-\phi_\lambda^2)\,d\lambda.
\end{equation}
The asymptotics (\ref{Agrinf}) and (\ref{d}) imply that the first sum in (\ref{x.1}) is $o(1)$
as $\rho\to 0$. The second sum coincides with
\begin{equation}\label{x.2}
\sum_{n=1}^M \oint_{|x_n|=\rho^{1/r_n}}\Phi_n(x_n)\left[\frac{2R^{z, x_n}}
{r_n x_n^{2r_n-2}}+\frac{1}{x_n^{2r_n}}\left(1-\frac{1}{r_n^2}\right)\right]r_n x_n^{r_n-1}\,dx_n.
\end{equation}
Here we have used (\ref{Agrpoint}); the function $\Phi_n$ is from Corollary \ref{kratnost}.
Now using Corollary \ref{kratnost} together with Cauchy formula and taking the limit
$\rho\to 0$ we get (\ref{lem}).

 $\square$

The rest of the proof relies on the method proposed in \cite{ZT2}.
Denote by $\Sigma_2$ the second term in (\ref{osn}). Using  (\ref{QuaziAlfors}) and the equality
${F_m}_{\bar\lambda}=0$, we get the relation
\begin{equation}\label{y.1}
\frac{\partial}{\partial \lambda_m}|\phi_\lambda|^2=-(F_m|\phi_\lambda|^2)_\lambda
-({F_m}_\lambda\phi_{\bar\lambda})_\lambda=-(F_m|\phi_\lambda|^2)_\lambda-({F_m}_\lambda\phi_\lambda)_{\bar\lambda}
-({F_m}_\lambda\phi_{\bar\lambda})_\lambda.
\end{equation}
This gives
\ben
\Sigma_2=-\frac{i}{2}\left(\sum_{k=1}^N\oint_{\partial\Omega_\rho^{(k)}}
F_m|\phi_\lambda|^2\,d\bar\lambda-\oint_{\partial\Omega_\rho^{(k)}}{F_m}_\lambda\phi_\lambda\,d\lambda+
\oint_{\partial\Omega_\rho^{(k)}}{F_m}_\lambda\phi_{\bar\lambda}d\bar\lambda\right)=
\een
\ben
=-\frac{i}{2}\sum_{\lambda_j}\sum_{p=1}^{r_j}\left(\oint_{|\lambda^{(p)}-\lambda^{(p)}_j|=\rho}
F_m|\phi_\lambda|^2d\bar\lambda-\oint_{|\lambda^{(p)}-\lambda^{(p)}_j|=\rho}{F_m}_\lambda\phi_\lambda
d\lambda+\oint_{|\lambda^{(p)}-\lambda^{(p)}_j|=\rho}{F_m}_\lambda\phi_{\bar\lambda}d\bar\lambda\right)
\een
\begin{equation}\label{y.2}
-\frac{i}{2}
\sum_{k=1}^N\left(\oint_{|\lambda^{(k)}|=1/\rho}F_m|\phi_\lambda|^2d\bar\lambda-
\oint_{|\lambda^{(k)}|=1/\rho}{F_m}_\lambda\phi_\lambda\,d\lambda+
\oint_{|\lambda^{(k)}|=1/\rho}{F_m}_\lambda
\phi_{\bar\lambda}d\bar\lambda\right),
\end{equation}
Let
\ben
I_1^n(\rho)=\sum_{p=1}^{r_n}\oint_{|\lambda^{(p)}-\lambda_n^{(p)}|=\rho}F_m|\phi_\lambda|^2d \bar\lambda\;;
\;\;\;\;\ I_2^n(\rho)=\sum_{p=1}^{r_n}\oint_{|\lambda^{(p)}-\lambda_n^{(p)}|=\rho}
{F_m}_\lambda\phi_\lambda\,d\lambda\;;
\een
\ben
I_3^n(\rho)=\sum_{p=1}^{r_n}\oint_{|\lambda^{(p)}-\lambda_n^{(p)}|=
\rho}{F_m}_\lambda\phi_{\bar\lambda}d\bar\lambda\;.
\een

 We have
\ben
I_1^n(\rho)=
\delta_{nm}\sum_{p=1}^{r_n}\oint_{|\lambda^{(p)}-\lambda_n^{(p)}|=\rho}|\phi_\lambda|^2
d\bar\lambda
\een
\ben
+\oint_{|x_n|=\rho^{1/r_n}}\left[\frac{1}{(r_n-1)!}\Phi_n^{(r_n-1)}(0)x_n^{r_n-1}+
\frac{1}{r_n!}\Phi_n^{(r_n)}(0)x_n^{r_n}+O(|x_n|^{r_n+1}\right]
\een
\ben
\times\left(\frac{|\phi_{x_n}^{int}|^2}{r_n x_n^{r_n-1}{\bar x}_n^{r_n-1}}+
\frac{1-r_n}{r_n^2}\frac{\phi_{x_n}^{int}}{{\bar x}_n^{r_n}x_n^{r_n-1}}
+\frac{1-r_n}{r_n^2}\frac{\phi_{{\bar x}_n}^{int}}{{\bar x}_n^{r_n-1}x_n^{r_n}}
+\left(\frac{1}{r_n}-1\right)^2\frac{1}{x_n^{r_n}{\bar x}_n^{r_n}}\right)r_n{\bar
x}_n^{r_n-1}d{\bar x}_n
\een
\ben
=\delta_{nm}\sum_{p=1}^{r_n}\oint_{|\lambda^{(p)}-\lambda_n^{(p)}|=\rho}|\phi_\lambda|^2
d\bar\lambda+2\pi i\frac{(1/r_n-1)^2}{(r_n-1)!}\Phi_n^{(r_n)}(0)+2\pi i\frac{1-r_n}{r_n(r_n-1)!}
\Phi_n^{(r_n-1)}(0)\phi_{x_n}^{int}(0)+o(1)
\een
as $\rho\to 0$.

We get also
\ben
I_2^n(\rho)=\oint_{|x_n|=\rho^{1/r_n}}\left(\frac{1}{r_n{x_n}^{r_n-1}}\phi_{x_n}^{int}+
\left(\frac{1}{r_n}-1\right)\frac{1}{x_n^{r_n}}\right)\left(\frac{1}{(r_n-2)!}\Phi_n^{(r_n-1)}(0)x_n^{r_n-2}\right.
\een
\ben
\left.+
\frac{1}{(r_n-1)!}\Phi_n^{(r_n)}(0)x_n^{r_n-1}+O(|x_n|^{r_n})\right)dx_n
\een
\ben
=-2\pi i\left(\frac{1}{r_n}-1\right)\frac{1}{(r_n-1)!}\Phi_n^{(r_n)}(0)-2\pi i\frac{1}{r_n(r_n-2)!}
\phi_{x_n}^{int}(0)\Phi_n^{(r_n-1)}(0)+o(1)
\een
and
\ben
I^n_3(\rho)=\oint_{|x_n|=\rho^{1/r_n}}\left(\frac{1}{(r_n-2)!}\Phi_n^{(r_n-1)}(0)x_n^{r_n-2}+
\frac{1}{(r_n-1)!}\Phi_n^{(r_n)}(0)x_n^{r_n-1}+O(|x_n|^{r_n})\right)
\een
\ben
\times\left(\frac{1}{r_n{\bar x}_n^{r_n-1}}\phi^{int}_{{\bar x}_n}+(\frac{1}{r_n}-1)\frac{1}
{{\bar x}_n^{r_n}}\right)(\frac{{\bar x}_n}{x_n})^{r_n-1}d{\bar x}_n=
2\pi i\frac{(1/r_n-1)}{(r_n-1)!}\Phi_n^{(r_n)}(0)+o(1)\;.
\een

We note that
\ben
I_1^n-I_2^n+I_3^n=\delta_{nm}\sum_{p=1}^{r_n}\oint_{|\lambda^{(p)}-\lambda_n^{(p)}|=\rho}|\phi_\lambda|^2
d\bar\lambda+\frac{2\pi
i}{(r_n-1)!}\Phi_n^{(r_n)}(0)[(1/r_n-1)^2+2(1/r_n-1)]+o(1)
\een
\ben
=\delta_{nm}\sum_{p=1}^{r_n}\oint_{|\lambda^{(p)}-\lambda_n^{(p)}|=\rho}|\phi_\lambda|^2
d\bar\lambda-\frac{2\pi
i}{(r_n-1)!}\left(1-\frac{1}{r_n^2}\right)\Phi_n^{(r_n)}(0)+o(1)\;.
\een
It is easy to  verify that
\ben
\sum_{k=1}^N\Big(\oint_{|\lambda^{(k)}|=1/\rho}F_m|\phi_\lambda|^2d\bar\lambda-
\oint_{|\lambda^{(k)}|=1/\rho}{F_m}_\lambda\phi_\lambda\,d\lambda+\oint_{|\lambda^{(k)}|=1/\rho}{F_m}_\lambda
\phi_{\bar\lambda}d\bar\lambda\Big)=o(1)\;,
\een
so we get
\begin{equation}\label{y.3}
\Sigma_2=-\frac{i}{2}\left(\sum_{l=1}^{r_m}\oint_{|\lambda^{(l)}-\lambda_m^{(l)}|=\rho}
|\phi_\lambda|^2d\bar\lambda-2\pi i\sum_{n=1}^M\frac{1}{(r_n-1)!}
\left(1-\frac{1}{r_n^2}\right)\Phi_n^{(r_n)}(0)\right)+o(1)\;.
\end{equation}
Now Lemma 7, (\ref{osn}) and (\ref{y.3}) imply that
\begin{equation}\label{y.4}
\frac{\partial }{\partial \lambda_m}Q^\rho=
\frac{2\pi}{(r_m-2)!\,r_m}\left(\frac{d}{dx_m}\right)^{r_m-2}R^{z, x_m}|_{x_m=0}
+o(1).
\end{equation}
To prove Theorem 2 it is sufficient to observe that
the term $o(1)$ in (\ref{y.4}) is uniform with respect to parameters $(\lambda_1, \dots, \lambda_M)$ belonging to a compact neighborhood of the initial
point $(\lambda_1^0, \dots, \lambda_M^0)$.

$\square$

\begin{corollary} The formulas for functions ${\mathbb S}_{ell}$ and ${\mathbb S}_{rat}$ can be rewritten as follows:
\begin{equation}\label{prosttor}
{\mathbb S}_{ell}(\lambda_1, \dots, \lambda_M)=\sum_{m=1}^M\frac{r_m-1}{2}\phi^{int}(x_m, {\bar x}_m)
\big|_{x_m=0}-\sum_{k=1}^{N}
\phi^\infty(\infty^{(k)}),
\end{equation}
\begin{equation}\label{prostsph}
{\mathbb S}_{rat}(\lambda_1, \dots, \lambda_M)=
\sum_{m=1}^M\frac{r_m-1}{2}\phi^{int}(x_m, {\bar x}_m)\big|_{x_m=0}-\sum_{k=2}^{N}
\phi^\infty(\infty^{(k)}).
\end{equation}
Here $\infty^{(k)}$ is the infinity of the $k$-th sheet of covering (\ref{nakr});
$\phi^\infty(\infty^{(k)})=\phi^{\infty}(\zeta, \bar\zeta)\big|_{\zeta=0}$; $\zeta=1/\lambda$ is
 the  local
parameter near $\infty^{(k)}$.
\end{corollary}

{\bf Proof.} Using the Liouville equation (\ref{Li}), the Stokes theorem and the asymptotics from Lemma 2,
 we get in the  case $g=1$:
\ben
Q^{\rho}
=\sum_{k=1}^N\int\int_{\Omega_\rho^k}(\phi_\l\phi)_{\bar\l}-\phi_{\l\bar\l}\phi \,dS=
\frac{1}{2i}\sum_{k=1}^N\int_{\partial\Omega_\rho^k}\phi_\l\phi\,d\l=
\een
\ben
=\frac{1}{2i}\left(\sum_{m=1}^M\oint_{|x_m|=\rho^{1/r_m}}
\left\{\frac{1}{r_m}\phi^{int}_{x_m} x_m^{1-r_m}+(\frac{1}{r_m}-1)x_m^{-r_m}\right\}\left\{\phi^{int}
+2(1-r_m)\log|x_m|-2\log r_m\right\}r_m x_m^{r_m-1}\,dx_m\right.
\een
\ben
\left. +\sum_{k=1}^N\oint_{|\l|=1/\rho}\left\{-\phi^\infty_\zeta\l^{-2}-\frac{2}{\l}\right\}\left\{\phi^\infty-
4\log|\l|\right\}\,d\l\right)=-\pi\sum_{m=1}^M(1-r_m)\phi^{int}(x_m)\big|_{x_m=0}-
2\pi\sum_{k=1}^N\phi^\infty(\infty^{(k)})
\een
\ben
-\left(4N+\sum_{m=1}^M\frac{(r_m-1)^2}{r_m}\right)2\pi\log \rho
-2\pi\sum_{m=1}^{M}(1-r_m)\log r_m+o(1)\;,
\een
as $\rho\to 0$.
This implies (\ref{prosttor}).

In case $g=0$ we repeat the same calculation, omitting the integrals around the infinity of the first sheet.

$\square$

\subsection{Factorization of the Dirichlet integral and the tau-functions of rational and elliptic
coverings}

Now we are in a position to calculate the Bergmann tau-function itself.
For rational coverings the Wirtinger and Bergmann tau-functions trivially coincide, in the elliptic case
the expression for the Wirtinger tau-function follows from that for the Bergmann one.

We start with the tau-functions of elliptic coverings.
\begin{theorem}\label{TORR} In case $g=1$ the Bergmann
tau-function of the covering $\L$ is given by the following expression:

\begin{equation}\label{taufinal}
\tau_B=[{\theta_1}'(0\,|\,\mu)]^{2/3}\frac{\prod_{k=1}^{N} h_k^{1/6}}{\prod_{m=1}^M f_m^{(r_m-1)/12}}\ \,
,
\end{equation}
where $v(P)$ is the normalized abelian differential on the torus $\L$; $v(P)=f_m(x_m)dx_m$ as
$P\to P_m$ and  $f_m\equiv f_m(0)$; $v(P)=h_k(\zeta)d\zeta$ as
$P\to \infty^{(k)}$ and  $h_k\equiv h_k(0)$; $\mu$ is the
$b$-period of the differential $v(P)$.
\end{theorem}

{\bf Proof.} It is sufficient to observe that $$\phi^{int}(x_m, x_m)=\log
U'(x_m)+\log \overline{U'(x_m)}=\log|f_m(x_m)|^2$$ in a
neighborhood of $P_m$ and $$\phi^\infty(\zeta,
\zeta)=\log|h_k(\zeta)|^2$$in a neighborhood
 of $\infty^{(k)}$ and to make use of (\ref{prosttor}) and (\ref{Hurtor}).
$\square$

Now  Theorem \ref{TORR}, the link (\ref{svBergWirt}) between the Bergmann and Wirtinger tau-functions,
and the Jacobi formula $\theta_1'=\pi \theta_2\theta_3\theta_4$ imply the following corollary
\begin{corollary}
The Wirtinger tau-function of the elliptic covering $\L$ is given by the formula
\begin{equation}\label{otvrod1}
\tau_W=\frac{\prod_{k=1}^{N} h_k^{1/6}}{\prod_{m=1}^M f_m^{(r_m-1)/12}}\ \, .
\end{equation}
\end{corollary}
We notice that the result (\ref{otvrod1}) does not depend on
normalization of the  holomorphic differential $v(P)$:
if one makes a transformation $v(P)\to C v(P)$ with an arbitrary constant $C$, this constant cancels out in
(\ref{otvrod1})
due to the  Riemann-Hurwitz formula.

For the rational case the Bergmann and Wirtinger tau-functions coincide.

\begin{theorem}\label{tausphfin}
In case $g=0$ the tau-functions of the covering $\L$ can be
calculated by the formula
\begin{equation}\label{tsphf}
\tau_W\equiv\tau_B=
\frac
{\prod_{k=2}^N(\frac{dU}{d\zeta_k}\big|_{\zeta_k=0})^{1/6}}
{\prod_{m=1}^M(\frac{dU}{dx_m}\big|_{x_m=0})^{(r_m-1)/12}},
\end{equation}
where $x_m$ is the local parameter near the branch point $P_m$,
$\zeta_k$ is the local parameter near the infinity of the $k$-th
sheet. (We recall that the map $U$ is chosen in such a way that
$U(\infty^{(1)})=\infty$.)
\end{theorem}

The proof is essentially the same.

\begin{remark}\label{stein}{\rm The fractional powers at the right hand sides of formulas
(\ref{tsphf}) and (\ref{otvrod1}) are understood in the sense of the analytical continuation.
The arising monodromies are just the monodromies generated by the flat connection $d_W$.
It should be noted that  the $12$-th powers of tau-functions
(\ref{tsphf}) and (\ref{otvrod1}) are single-valued global holomorphic functions
on the Hurwitz space ${\cal U}(\L)$.
}
\end{remark}

It is instructive to illustrate  the  formulas (\ref{tsphf}) and
(\ref{taufinal}) for  the simplest two-fold
coverings with two ($g=0$) and four ($g=1$) branch points.

\subsubsection{Tau-function of a two-fold rational covering}
 Consider the  covering of  ${\mathbb  P}^1$ with  two sheets  and two
  branch points $\l_1$ and $\l_2$.
Then $g=0$ and
\begin{equation}\label{primer}
U(\lambda)=\frac{1}{2}\left(\lambda+\frac{\lambda_1+\lambda_2}{2}+
\sqrt{(\lambda-\lambda_1)(\lambda-\lambda_2)}\right).
\end{equation}
 We get
\ben
\{U(x_1),
x_1\}_{x_1=0}=\left\{x_1^2+x_1\sqrt{\l_1-\l_2+x_1^2},
x_1\right\}\Big|_{x_1=0}
\een
\ben
=\left\{\sqrt{\l_1-\l_2}x_1+x_1^2+\frac{x_1^3}
{2\sqrt{\l_1-\l_2}}, x_1\right\}\Big|_{x_1=0}=\frac{3}{\l_2-\l_1}
\een
 and
\begin{equation}\label{newone}
\{U(x_2), x_2\}|_{x_2=0}=\f{3}{\l_1-\l_2}\;.
\end{equation}
 Now direct integration of equations (\ref{newone})  gives the following result:
\begin{equation}
\tau_W=\tau_B=\;(\l_1-\l_2)^{1/4}\;
\la{tau000}
\end{equation}
(up to a multiplicative constant).
 On the other hand, to apply the general formula (\ref{tsphf}), we find
\ben
U_{x_1}(0)=\frac{1}{2}\sqrt{\l_1-\l_2}\;;\hskip0.9cm
U_{x_2}(0)=\frac{1}{2}\sqrt{\l_2-\l_1}\;,
\een
\ben
U(\zeta_2)=\frac{1}{2}\left(\frac{1}{\zeta_2}+\frac{\l_1+\l_2}{2}-\frac{1}{\zeta_2}
\sqrt{(1-\zeta_2\l_1)(1-\zeta_2\l_2)}\right)
\een
\ben
=\frac{\l_1+\l_2}{2}+\frac{(\l_1-\l_2)^2}{16}\zeta_2+\dots\;.
\een
Therefore, our formula (\ref{tsphf}) in this case also gives rise to (\ref{tau000}).

\subsubsection{Tau-functions of two-fold elliptic coverings}

 Consider the two-fold covering $\L$ with four branch points:
\begin{equation}
\mu^2=(\l-\l_1)(\l-\l_2)(\l-\l_3)(\l-\l_4)\ .
\la{eqtor}
\end{equation}
There are two ways to compute the tau-function on the space of such
coverings. On one hand, since the elliptic curve $\L$  belongs to the  hyperelliptic class, we can
apply known formula (\ref{Btauhe}) which gives:
\begin{equation}
\tau_B(\l_1, \dots, \l_{4})={\cal A}\prod_{m,n=1, \dots 4; m<n}
(\l_m-\l_n)^{1/4},
\la{tauKK}
\end{equation}
where ${\cal A}=\oint_a \f{d\l}{\mu}$ is the $a$-period of the non-normalized
holomorphic differential.

On the other hand, to apply the formula (\ref{taufinal}) to this case,
we notice that the normalized holomorphic differential on $\L$ is
equal to
\ben
v(P)=\f{1}{{\cal A}}\f{d\l}{\mu}\;;
\een
the local parameters near $P_n$ are $x_n=\sqrt{\l-\l_n}$.
Therefore,
\ben
f_m=2{\cal A}^{-1}\prod_{n\neq m}(\l_m-\l_n)^{-1/2}\;,\hskip0.8cm
h_k=(-1)^k{\cal A}^{-1}\;,\hskip0.8cm  k=1, 2\;.
\een
According to the Jacobi formula $\theta_1'=\pi \theta_2\theta_3\theta_4$;
moreover, the  genus 1 version of Thomae formulas for
theta-constants gives
$\theta_k^4 =\pm\f{{\cal A}^2}{(2\pi
i)^2}(\l_{j_1}-\l_{j_2})(\l_{j_3}-\l_{j_4})$, where $k=2,3,4$ and $(j_1,\dots, j_4)$
are appropriate permutations of $(1,\dots,4)$.
Computing $\theta_1'$ according to these expressions, we again get
(\ref{tauKK}).

\subsection{The Wirtinger tau-function  and isomonodromic deformations}

In \cite{Koro01} it was given a solution to  a class
of the Riemann-Hilbert problems with quasi-permutation monodromies in
terms of Szeg\"o kernels on
branched coverings of $\CP1$. The isomonodromic tau-function of Jimbo
and Miwa associated to these Riemann-Hilbert problems is closely
related to the tau-functions of the branched coverings considered in this paper.

Here we briefly outline this link  for the  genus zero coverings $\L$.
So, let $\L$ be biholomorphically equivalent to the Riemann sphere ${\mathbb P}^1$
with global coordinate $z$.
Introduce the ``prime-forms''  on the $z$-sphere and the $\l$-sphere:
\begin{equation}
E(z,z_0)=\f{z-z_0}{\sqrt{dz}\sqrt{dz_0}}\;,\hskip0.8cm
E_0(\l,\l_0)=\f{\l-\l_0}{\sqrt{d\l}\sqrt{d\l_0}}\;.
\la{prime}
\end{equation}
Define a $N\times N$ matrix-valued function $\Psi(\l,\l_0)$ for $\l$
belonging to a small neighborhood of $\l_0$:
\begin{equation}
\Psi_{jk}(\l,\l_0)=\f{E_0(\l,\l_0)}{E(\l^{(k)},\l_0^{(j)})}=
\f{(\l-\l_0)\sqrt{z'(\l^{(k)})}\sqrt{z'(\l_0^{(j)})}}{z(\l^{(k)})-z(\l_0^{(j)})}\;,
\la{psidef}
\end{equation}
where $z'= dz/d\l$.
To compute the determinant of the matrix $\Psi$ we use  the following identity
 for two arbitrary sets of complex numbers $z_1,\dots,z_N,\mu_1,\dots,\mu_N$:
\begin{equation} \la{fayform}
\det_{N\times N}\left\{\f{1}{z_j-\mu_k}\right\}=
\f{\prod_{j< k}(z_j-z_k)(\mu_k-\mu_j)}{\prod_{j,k}(z_j-\mu_k)}\;.
\end{equation}
Using this relation, we find that
\ben
\det\Psi=(\l-\l_0)^N\prod_{k=1}^N \{z_\l(\l^{(k)})z_\l(\l_0^{(k)})\}^{N/2}
\f{\prod_{j<k}\{z(\l^{(k)})-z(\l^{(j)})\}\{z(\l_0^{(j)})-z(\l_0^{(k)})\}}
{\prod_{j,k}\{z(\l^{(k)})-z(\l_0^{(j)})\}}\;\;.
\een
This expression is symmetric with respect to interchanging of any two sheets, therefore, it is
a single-valued function of $\l$ and $\l_0$. Moreover, it is
non-singular (and equal to  $1$) as $\l=\l_0$,
and non-singular as $\l\to\infty$. Therefore, it is globally non-singular, thus
 identically equal to  $1$.

The function $\Psi$ obviously equals to the unit matrix as $\l\to\l_0$.
The only singularities of the function $\Psi$  in $\l$-plane are the branch points $\l_m$.
These are regular singularities with quasi-permutation  monodromy matrices with non-vanishing entries
equal to $\pm 1$.

Therefore,  function $\Psi(\l)$, being analytically continued from a small neighborhood of point $\l_0$ to the
universal covering of $\CP1\setminus\{\l_1,\dots,\l_m\}$, gives  a
solution to the  Riemann-Hilbert problem with regular singularities at
the points $\l_m$ and quasi-permutation monodromy matrices. It is non-degenerate outside of $\{\l_m\}$,
equals $I$ at $\l=\l_0$, and satisfies the equations
\begin{equation}
\f{\p\Psi}{\p\l}=\sum_{m=1}^M \f{A_m}{\l-\l_m}\Psi\;,\hskip0.8cm
\f{\p\Psi}{\p\l_m}=-\f{A_m}{\l-\l_m}\Psi
\la{lslambda}
\end{equation}
for some $N\times N$ matrices $\{A_m\}$ depending on $\{\l_m\}$.
Compatibility of equations (\ref{lslambda}) implies
the  Schlesinger system for the functions $A_m(\{\l_n\})$. The corresponding Jimbo-Miwa
tau-function
$\tau_{JM}(\{\l_m\})$ is  defined by the equations
\begin{equation}
\f{\p\log\tau_{JM}}{\p\l_m}=\f{1}{2}\res|_{\l=\l_m}\tr(\Psi_\l\Psi^{-1})^2\;.
\la{deftau0}
\end{equation}
The tau-function, as well as the expression $\tr(\Psi_\l\Psi^{-1})^2$, is
 independent of the normalization point $\l_0$; taking the limit $\l_0\to\l$ in this expression,
we get
\begin{equation}
\Psi_{jk}= \f{z_\l(\l^{(j)})z_\l(\l^{(k)})}{z(\l^{(j)})-z(\l^{(k)})}(\l_0-\l)+ O((\l-\l_0)^2)\;,\hskip1.0cm
\Psi_{jj}= 1+ o(1)\hskip0.4cm {\rm as}\hskip0.4cm \l_0\to\l
\end{equation}
and
\begin{equation}
\f{1}{2}\tr\left(\Psi_\l\Psi^{-1}(\l)\right)^2
=-\f{1}{(d\l)^2}\sum_{j\neq k} B\left(z(\l^{(j)}),z(\l^{(k)})\right)\;,
\la{trA2}
\end{equation}
where
\ben
B(z,\tilde{z})= \f{dz d\tilde{z}}{(z- \tilde{z})^2}
\een
is the Bergmann kernel on $\CP1$.
Consider the behavior of expression (\ref{trA2}) as $\l\to\l_m$;
suppose that the sheets glued at the ramification point  $P_m$ have numbers $s$ and $t$.
Then, since $d\l=2x_m dx_m$, we have as $\l\to\l_m$,
\ben
\f{1}{2}\tr(\Psi_\l\Psi^{-1}(\l))^2=-\f{1}{4(\l-\l_m)}\f{z_{x_m}(\l^{(s)})z_{x_m}(\l^{(t)})}
{\left[z(\l^{(s)})-z(\l^{(t)})\right]^2}+ O(1)
\een
\ben
=
-\f{1}{4(\l-\l_m)}\left(\f{1}{[x_m(\l^{(s)})-x_m(\l^{(t)})]^2}+\f{1}{6}\{z,x_m\}|_{x_m=0}\right)+
O(1)
\een
\ben
= -\f{1}{4(\l-\l_m)}\left(\f{1}{4(\l-\l_m)}+\f{1}{6}\{z,x_m\}|_{x_m=0}\right)+O(1)\;.
\een

Therefore, the definition of isomonodromic tau-function  (\ref{deftau0}) gives rise to
\begin{equation}
\f{\p\log\tau_{JM}}{\p\l_m}=-\f{1}{24}\{z,x_m\}|_{x_m=0}\;;
\la{deftau00}
\end{equation}
thus, in genus zero we get the following relation between isomonodromic and Wirtinger tau-functions:
$$
\tau_{JM}=\{{\tau_W}\}^{-1/2}\;,
$$
where $\tau_W$ is given by (\ref{tsphf}).

\section{The case of higher genus}\label{SHG}
In this section we calculate the modulus square of the Bergmann
and Wirtinger tau-functions for an arbitrary covering of genus $g>1$.

Let $\L_0$ be a point of
$\hat{\cal U}(\L)$. In a  small
neighborhood of $\L_0$ we may consider
the branch points  $\l_1, \dots, \l_M$ as local coordinates on $\hat{\cal U}(\L)$.

The tau-function $\tau_B$ (a section of the Bergmann line bundle)
can be considered as a holomorphic function in this small neighborhood
of $\L_0$. Its modulus
square, $|\tau_B|^2$
is the restriction of a section of the ``real'' line bundle ${\cal T_B}\otimes\overline{{\cal T_B}}$.

To compute $|\tau_B|^2$ we are to find a real-valued potential $\log|\tau_B|^2$
such that
\begin{equation}\label{potB}
\frac{\partial \log|\tau_B|^2}{\partial \l_m}={\cal B}_m\;; \ \ m=1, \dots, M\;.
\end{equation}

If the covering $\L$ has genus $g>1$ then it
is biholomorphically equivalent to the quotient space ${\mathbb H}/\Gamma$, where
${\mathbb H}=\{z\in {\mathbb C}\,:\, \Im z>0\}$; $\Gamma$ is a strictly hyperbolic  Fuchsian group.
Denote by $\pi_F:{\mathbb H}\rightarrow \L$ the natural projection. The Fuchsian projective connection
on $\L$ is given by the Schwarzian derivative $\{z, x\}$,
where $x$ is a local coordinate of a point $P\in \L$, $z\in {\mathbb H}$, $\pi_F(z)=P$.

We recall the variational formula  (\cite{ZT4}, see also \cite{Fay})) for the determinant of the Laplacian
on the Riemann surface $\L$:
$$\delta_\mu\log\left(\frac{{\rm det}\Delta}{{\rm det}\Im {\mathbb B}}\right)=-\frac{1}{12\pi i}
\int_\L(S_B-S_F)\mu,$$
where ${\mathbb B}$ is the matrix of $b$-periods,
$S_B$ is the Bergmann projective connection, $S_F$ is the Fuchsian projective connection,
$\mu$ is a Beltrami differential.
Since, as we discussed above, the derivation with respect to $\l_m$ corresponds to the Beltrami differential
$\mu_m$ from (\ref{sch2}), we conclude that
\begin{equation}\label{detL}
-\frac{1}{6r_m\,(r_m-2)!}\left(\frac{d}{dx_m}\right)^{r_m-2}\left(S_B(x_m)-\{z, x_m\}\right)\big|_{x_m=0}=
\frac{\partial}{\partial\l_m}\log\left(\frac{{\rm det}\Delta}{\text{det}\,\Im {\mathbb B}}\right)
\end{equation}
\begin{remark}\rm{This formula explains the appearance of the factor $-\frac{1}{6}$ in the definition
(\ref{Hurw}) of the connection coefficient ${\cal B}_m$.}
\end{remark}

Therefore, the calculation of the modulus of the Bergmann
tau-function of the covering $\L$ reduces to the problem of finding
a real-valued function ${\mathbb S}_{Fuchs}(\l_1, \dots, \l_M)$ such that
\begin{equation}\label{uravfuchs}
\frac{\partial{\mathbb S}_{Fuchs}}{\partial\l_m}=\frac{1}{r_m\,(r_m-2)!}\left(\frac{d}{dx_m}\right)^{r_m-2}
\{z, x_m\}\Big|_{x_m=0},\ \ \ \ m=1, \dots M.
\end{equation}

Another link of $|\tau_B|^2$ with known objects can be established if we introduce the Schottky uniformization
of the covering $\L$.
Namely, the covering $\L$ (of genus $g>1$) is biholomorphically equivalent to the quotient space
$$\L={\bf D}/\Sigma\ \, ,$$
where $\Sigma$ is a (normalized) Schottky group, ${\bf D}\subset {\mathbb P}^1$ is its region of discontinuity.
Denote by $\pi_\Sigma:{\bf D}\rightarrow\L$  the natural projection.

Introduce the Schottky projective connection on $\L$ given  by the Schwarzian derivative
$\{\omega, x\}$,
where $x$ is a local coordinate of a point $P\in \L$; $\omega\in{\bf D}$;
$\pi_\Sigma(\omega)=P$.

Due to the formula (\ref{osnbelt}) and the results of \cite{Zograf}
(namely, see Remark 3.5 in \cite{Zograf}),
we have
\begin{equation}\label{dbar}
-\frac{1}{6r_m(r_m-2)!}\left(\frac{d}{dx_m}\right)^{r_m-2}\left(S_B(x_m)-\{\omega, x_m\}\right)\Big|_{x_m=0}=
\frac{\partial}{\partial\l_m}\log |\text{det}\bar\partial|^2.
\end{equation}
Here $\text{det}\,\bar\partial$ is the holomorphic determinant of the family of $\bar\partial$-operators
(this holomorphic determinant can be
considered as a
nowhere vanishing holomorphic function on the Schottky space;
see Theorem 3.4 \cite{Zograf}
for precise definitions and an explicit formula for $|\text{det}\bar\partial|^2$).

Therefore, the calculation of the modulus square of the Bergmann tau-function of the covering $\L$ reduces
to the integration of the following system of equations for real-valued function ${\mathbb S}_{Schottky}$:
\begin{equation}\label{uravsch}
\frac{\partial{\mathbb S}_{Schottky}}{\partial\l_m}=
\frac{1}{r_m(r_m-2)!}\left(\frac{d}{dx_m}\right)^{r_m-2}\{\omega, x_m\}\Big|_{x_m=0},\ \ \ \ m=1, \dots M.
\end{equation}

In the following two subsections we solve, first, system
(\ref{uravsch}) and, second, system (\ref{uravfuchs}).

\subsection{The Dirichlet integral and the Schottky uniformization}
\subsubsection{The Schottky uniformization and the flat metric on dissected Riemann surface}
{\bf The Schottky uniformization.}
We refer the reader to \cite{ZT3} for a brief review of  Schottky groups and the Schottky uniformization
theorem.

Fix some marking of the Riemann surface $\L$ (i. e. a point $x_0$ in $\L$ and some system of generators
$\a_1, \dots, \a_g, \b_1, \dots, \b_g$ of the fundamental group $\pi_1(\L, x_0)$ such that
$\Pi_{i=1}^g\a_i^{-1}\beta_i^{-1}\a_i\b_i=1$).

The marked surface $\L$ is biholomorphically equivalent to the quotient space ${\bf D}/\Sigma$,
where $\Sigma$ is a normalized marked Schottky group, ${\bf D}\subset {\mathbb P}^1$ is its region of discontinuity.
(A Schottky group is said to be marked  if a relation-free system of generators $L_1, \dots,
L_g$ is chosen in it. For the normalized Schottky group $L_1(\omega)=k_1\omega$ with $0<|k_1|<1$ and
the attracting fixed point of the transformation $L_2$ is $1$.)

Choose a fundamental region $D_0$ for $\Sigma$ in ${\bf D}$. This is a region in ${\mathbb P}^1$ bounded by
$2g$ disjoint Jordan curves $c_1, \dots, c_g, c_1', \dots, c_g'$ with $c_i'=-L_i(c_i), \, i=1, \dots, g$;
the curves $c_i$ and $c_i'$ are oriented as the components of  $\partial D_0$, the minus sign means the reverse
orientation.

Let $\pi_\Sigma:{\bf D}\rightarrow \L$ be the natural projection. Set $C_i=\pi_\Sigma(c_i)$.

Denote by $\Ld$ the dissected surface $\L\setminus \cup_{i=1}^gC_i$. The map $\pi_\Sigma:D_0\rightarrow
\Ld$ is invertible; denote the inverse map by $\O_0$.
\subsubsection{The flat metric on $\Ld$}
Let $x$ be a local parameter on $\Ld$. Define a flat metric $e^{\phi(x, \bar x)}|dx|^2$ on $\Ld$
by
\begin{equation}\label{metric}
e^{\phi(x, \bar x)}|dx|^2=|d\omega|^2.
\end{equation}
Here $\omega\in D_0$, $\pi_\Sigma(\omega)=x$. Thus, to each local chart with local
parameter $x$ there corresponds a function $\phi(x, \bar x)$.
 We specify the function $\phi^{ext}(\l, \bar\l)$ of local parameter $\l$ by
\begin{equation}\label{defmetric}
e^{\phi^{ext}(\l, \bar\l)}|d\l|^2=|d\omega|^2=|\O_0'(\l)|^2|d\l|^2.
\end{equation}
Here $\omega\in {\bf D}$, $\pi_\Sigma(\omega)=P\in \L$ and $p(P)=\l$.

Introduce also the functions $\phi^{int}(x_m, \bar x_m)$, $m=1, \dots, M$ and $\phi^\infty(\zeta_k, \bar\zeta_k)$,
$k=1, \dots, N$
corresponding to the local parameters $x_m$ near the ramification points $P_m$ and the local parameters $\zeta_k=1/\l$
near the infinity of the $k$-th sheet. In the intersections of the local charts we have
\begin{equation}\label{tv}
e^{\phi^{int}(\lp, \bar x_m)}|d\lp|^2=e^{\phi^{ext}(\l, \bar\l)}|d\l|^2
\end{equation}
 and
\begin{equation}\label{besk}
e^{\phi^\infty(\zeta_k,
\bar\zeta_k)}|d\zeta_k|^2=e^{\phi^{ext}(\l, \bar\l)}|d\l|^2.
\end{equation}

Choose an element $L\in\Sigma$ and consider the fundamental region $D_1=L(D_0)$. Introduce the map
$\Omega_1:\Ld\rightarrow D_1$ and the metric $e^{\phi_1(x, x)}|dx|^2$ on $\Ld$ corresponding
to this new choice of fundamental region.

Since $\O_1(x)=L(\O_0(x))$, we have
\begin{equation}\label{zam1}
\phi_1(x, \bar x)=\phi(x, \bar x)+\log|L'(\O_0(x))|^2,
\end{equation}
\begin{equation}\label{zam2}
[\phi_1(x, \bar x)]_x=\phi_x(x, \bar x)+\frac{L''(\O_0(x))}{L'(\O_0(x))}\O_0'(x)
\end{equation}
and
\begin{equation}\label{zam3}
[\phi_1(x, \bar x)]_{\bar x}=\phi_{\bar x}(x, \bar x)+\frac
{\overline{L''(\O_0(x))}}{\overline{L'(\O_0(x))}}\overline{\O_0'(x)}.
\end{equation}

The following statements are complete analogs of those from section 3.1. Lemmas \ref{as} and
\ref{shwartz} are evident, to get lemmas \ref{diff}, \ref{funR} and Corollary \ref{kratnost1}
one only needs to change the map $U\,:\,\L\ni x\mapsto z\in\tilde\L$ to the map
$\Omega_0\,:\,\Ld\ni x\mapsto \omega\in D_0$ in the proofs of corresponding statements from
section 3.1. Since the map $\Omega_0$, similarly to the map $U$, depends on the branch points
$\l_1, \dots, \l_M$ holomorphically, all the arguments from \S 3.1 can be applied in the present context.

\begin{lemma}\label{as}
The derivative of the function $\phi^{ext}$ has the following asymptotics near the branch points and the
infinities of the sheets:
\begin{enumerate}
\item $|\phi_\lambda^{ext}(\lambda, \lambda)|^2=(\frac{1}{r_m}-1)^2|\lambda-\l_m|^{-2}+O(|\lambda-\l_m|^{-2+1/r_m
})$
as $\lambda\rightarrow \l_m$,
\item $|\phi_\lambda^{ext}(\lambda, \lambda)|^2=4|\lambda|^{-2}+O(|\lambda|^{-3})$ as $\lambda\rightarrow \infty$.
\end{enumerate}
\end{lemma}

Let $x$ be a local coordinate on $\L$. Set $R^{\omega, x}=\{\omega, x\}$, where $\omega\in{\bf D}$,
$\pi_\Sigma(\omega)=x$.

\begin{lemma}\label{shwartz}
\begin{enumerate}
\item
The Schwarzian derivative can be expressed as follows in terms of the function $\phi$ from
 (\ref{metric}):
\begin{equation}\label{Agregat1}
R^{\omega, x}=\phi_{xx}-\frac{1}{2}\phi_x^2.
\end{equation}
\item
In a neighborhood of a branch point $P_m$  there is the following relation between
 Schwarzian derivatives computed with respect to coordinates $\l$ and $x_m$:
\begin{equation}\label{Agrpoint1}
R^{\omega, \lambda}=\frac{1}{r_m^2}(\lambda-\l_m)^{2/r_m-2}R^{\omega, \lp}+
\left(\frac{1}{2}-\frac{1}{2r_m^2}\right)(\lambda-
\l_m)^{-2}.
\end{equation}
\item
Let $\zeta$ be the coordinate in a neighborhood of the infinity of any sheet of covering $\L$, $\zeta=1/\lambda$.
Then
 \begin{equation}\label{Agrinf1}
R^{\omega, \lambda}=\frac{R^{\omega, \zeta}}{\l^4}=O(|\lambda|^{-4}).
\end{equation}
\end{enumerate}
\end{lemma}

\begin{lemma}\label{diff}
The derivatives of the function $\phi$ with respect to $\l$ are related to
its derivatives with respect to the branch points as follows:
\begin{equation}\label{QuaziAlfors1}
\frac{\partial\phi}{\partial\lambda_m}+F_m\frac{\partial \phi}{\partial \lambda}+
\frac{\partial F_m}{\partial\lambda}=0\;,
\end{equation}
where
\begin{equation}\label{koeff1}
F_m=-\frac{[\O_0]_{\lambda_m}}{[\O_0]_\lambda}\;.
\end{equation}
\end{lemma}

\begin{lemma}\label{funR}
Denote the composition
$p\circ \pi_\Sigma$ by $R$. Then
\begin{enumerate}
\item
The following relation holds:
\begin{equation}\label{w1}
F_m=\frac{\partial R}{\partial \lambda_m}.
\end{equation}

\item
In a neighborhood of the point $\l_l$ the following
 asymptotics holds:
\begin{equation}\label{c1}
F_m=\delta_{lm}+o(1),
\end{equation}
where $\delta_{lm}$ is the Kronecker symbol.
\item
At the infinity of each sheet
the following asymptotics holds:
\begin{equation}\label{d1}
F_m(\lambda)=O(|\lambda|^2).
\end{equation}
\end{enumerate}
\end{lemma}

\begin{corollary}\label{kratnost1} Keep $m$ fixed and define $\Phi_n(x_n)\equiv F_m(\lambda_n+x_n^{r_n})$.
Then
\ben
\Phi_n(0)=\delta_{nm};\ \ \ \ \ \
\left(\frac{d}{dx_n}\right)^k\Phi_n(0)=0,\ \ \  k=1, \dots, r_n-2\;.
\een
\end{corollary}

\subsubsection{The regularized Dirichlet integral}

Assume that the ramification points and the infinities of sheets do not belong to the cuts $C_i$.

To the $k$-th sheet $\Ld^{(k)}$ of the dissected surface $\L$ (we should add some cuts connecting
the branch points)
 there corresponds
 the function $\phi_k^{ext}: \Ld^{(k)}\to {\mathbb R}$ which
is smooth in any domain $\Delta^k_r$
of the form $\Delta^k_\rho=\{\lambda\in \Ld^{(k)}: \forall m \;
|\lambda-\lambda_m|>\rho\;\; \& \;\;  |\lambda|<1/\rho\}$, where $\rho>0$
 and $\lambda_m$ are all
the branch points from the $k$th sheet $\Ld^{(k)}$ of $\Ld$.

The function $\phi_k^{ext}$ has finite limits at the cuts (except the endpoints which are the ramification points);
at the ramification points and at the infinity it possesses the asymptotics listed in Lemma \ref{as}.

Introduce the regularized Dirichlet integral
$$\int_{\Ld}|\phi_\l|^2\, dS.$$
Namely, set
\begin{equation}\label{action1}
Q_\rho=\sum_{k=1}^N\int_{\Delta_\rho^k}|\partial_\lambda\phi^{ext}_k|^2
\,dS\;,
\end{equation}
where $dS$ is the area element on ${\mathbb C}^1$:
$dS=\frac{|d\l\wedge d\bar\l|}{2}$.

According to Lemma 3 there exists the finite limit
\begin{equation}\label{reg1}
{\rm reg}\int_{\Ld}|\phi_\l|^2\, dS
=\lim_{\rho\to 0}
\left(Q_\rho+(4N+\sum_{m=1}^M\frac{(r_m-1)^2}{r_m})2\pi\log \rho\right)
\end{equation}

Now set
\begin{equation}\label{Main}
{\mathbb S}_{Schottky}(\l_1, \dots, \l_M)=
\frac{1}{2\pi}{\rm reg}\int_{\Ld}|\phi_\l|^2\, dS+\frac{i}{4\pi}\sum_{k=2}^g
\left\{
\int_{C_k}\phi(\l, \bar\l)
\frac{\overline{L_k''(\Omega_0(\l))}}{\overline{L_k'(\Omega_0(\l))}}\overline{\Omega_0'(\l)}\,d\bar\l-
\right.
\end{equation}
\begin{equation*}
\left.-\int_{C_k}\phi(\l, \bar\l)\frac{L_k''(\Omega_0(\l))}{L_k'(\Omega_0(\l))}\Omega_0'(\l)\,d\l+
\int_{C_k}\log|L_k'(\Omega_0(\l)|^2\frac{\overline{L_k''(\Omega_0(\l))}}{\overline{L_k'(\Omega_0(\l))}}
\overline{\Omega_0'(\l)}\,d\bar\l\right\}+2\sum_{k=2}^g\ln|l_k|^2.
\end{equation*}
Here $L_k$ are generators of the Schottky group $\Sigma$, the orientation of contours $C_k$ is defined
by the orientation of countours $c_k$ and the relations $C_k=\pi_\Sigma(c_k)$; the value of the function
$\phi(\l, \bar\l)$ at the point $\l\in C_k$ is defined as the limit
$\lim_{\mu\rightarrow\l}\phi(\mu, \bar\mu)$, $\mu=\pi_\Sigma(\omega)$ and $\omega$ tends to the contour
$c_k$ from the interior of the region $D_0$; $l_k$ is the left-hand lower element in the matrix representation
of the transformation $L_k\in{\rm PSL}(2, {\mathbb C})$. The summations at the right hand side
of (\ref{Main})  start from $k=2$ due to the normalization condition for the group $\Sigma$ (the terms with
$k=1$ are equal to zero).

Observe that the expression at the right hand side of (\ref{Main}) is real and does not depend on small
movings of the cuts $C_k$ (i. e. on a specific choice of the fundamental region $D_0$). In particular,
we can assume that the contours $C_k$ are $\{\l_1, \dots, \l_M\}$-independent.
(To see this one should make a simple calculation based on (\ref{zam2}), (\ref{zam3}) and the Stokes theorem.)
Thus all terms in this expression except the last one are rather natural. The role of the last term will
become clear later.

The main result of this section is the following theorem.
\begin{theorem}\label{sch}
For any $m=1, \dots, M$ the following equality holds
\begin{equation}\label{tsch}
\frac{\partial{\mathbb S}_{Schottky}(\l_1, \dots, \l_M)}{\partial \l_m}=
\frac{1}{(r_m-2)!\,r_m}\left(\frac{d}{dx_m}\right)^{r_m-2}R^{\omega, x_m}\Big|_{x_m=0}.
\end{equation}
\end{theorem}

\begin{remark} {\rm This result seems to be very similar to Theorem 1 from \cite{ZT3}. However, we would like to
emphasize that in oppose to \cite{ZT3} we deal here with the Dirichlet integral corresponding to
a {\it flat} metric. Thus, the following proof does not explicitly use
the Teichm\"uller theory and, therefore, is
more elementary than the proof of an analogous result in \cite{ZT3}.}
\end{remark}

{\bf Proof.}
Set
\begin{equation}\label{Sr}
S_\rho=Q_\rho+\frac{i}{2}\sum_{k=2}^g
\left\{
\int_{C_k}\phi(\l, \bar\l)
\frac{\overline{L_k''(\Omega_0(\l))}}{\overline{L_k'(\Omega_0(\l))}}\overline{\Omega_0'(\l)}\,d\bar\l-
\int_{C_k}\phi(\l, \bar\l)\frac{L_k''(\Omega_0(\l))}{L_k'(\Omega_0(\l))}\Omega_0'(\l)\,d\l+\right.
\end{equation}
\begin{equation*}\left.
+\int_{C_k}\log|L_k'(\Omega_0(\l)|^2\frac{\overline{L_k''(\Omega_0(\l))}}{\overline{L_k'(\Omega_0(\l)}}
\overline{\Omega_0'(\l))}\,d\bar\l\right\}.
\end{equation*}

 We recall that the contours $C_k$ are assumed to be $\{\l_1, \dots, \l_M\}$-independent.
From now on we write $\Omega(\l)$ and $\phi$ instead of $\Omega_0(\l)$ and $\phi^{ext}$.
Since $\phi_{\l\bar\l}=0$, we have $|\phi_\l|^2=(\phi_\l\phi)_{\bar\l}$. The Stokes theorem
and the formulas (\ref{zam1}), (\ref{zam2}) give
\begin{equation}\label{str1}
Q_\rho= -\frac{i}{2}\left[\sum_{n=1}^{M}\sum_{l=1}^{r_n}\oint_{|\l^{(l)}-\l_n|=\rho}\phi_\l\phi\,d\l
+\sum_{k=1}^N\oint_{|\l^{(k)}|=1/\rho}\phi_\l\phi\,d\l\right]-
\end{equation}
\begin{equation*}
-\frac{i}{2}
\sum_{k=2}^g\int_{C_k}\Big\{\phi_\l\phi-
\big[\phi_\l+\frac{L_k''(\Omega(\l))}{L_k'(\Omega(\l))}\Omega'(\l)\big]\big[\phi+\log|L_k'(\Omega(\l))|^2
\big]\Big\}\,d\l\ \,.
\end{equation*}
Here $\l^{(k)}$ denotes the point on the $k$-th sheet of the covering $\L$ whose projection to ${\mathbb P}^1$
is $\l$.

Denote the first term in (\ref{str1}) by $-\frac{i}{2}\big[T_\rho\big]$.
Substituting (\ref{str1}) into (\ref{Sr}) and using  the equalities
$\int_{C_k}d[\phi(\l, \bar\l)\log|L'(\O(\l))|^2]=0$ and $\int_{C_k}d[\log^2|L'_k(\O(\l))|^2]=0$, we get
\begin{equation}\label{str2}
S_\rho=-\frac{i}{2}\Big[T_\rho\Big]-\frac{i}{2}\sum_{k=2}^g\int_{C_k}
\phi_{\bar\l}(\l, \bar\l)\log|L_k'(\Omega(\l))|^2\,d\bar\l-\frac{i}{2}\sum_{k=2}^g\int_{C_k}\phi(\l, \bar\l)
\frac{L_k''(\Omega(\l))}{L_k'(\Omega(\l))}\Omega'(\l)\,d\l.
\end{equation}
\begin{lemma}\label{standchlen} For the first term in (\ref{str2}) we have the asymptotics
\begin{equation}\label{stch}
-\frac{i}{2}\frac{\partial}{\partial\l_m}\big[T_\rho\big]=\frac{2\pi}{(r_m-2)!\,r_m}\left
(\frac{d}{dx_m}\right)^{r_m-2}R^{\omega, x_m}\Big|_{x_m=0}+
\end{equation}
\begin{equation*}
+\frac{i}{2}\sum_{k=1}^g\int_{C_k\cup C_k^-}\left\{F_m(2\phi_{\l\l}-\phi_\l^2)+[F_m]_\l\phi_\l\right\}\,d\l
+o(1),
\end{equation*}
as $\rho\rightarrow 0$. Here $C_k^-$ is the contour $C_k$ provided by the reverse orientation,
the value of the integrand at a point $\l\in C_k^-$ is understood as the limit as $\mu\rightarrow\l$,
where $\mu=\pi_\Sigma(\omega)$, $\omega$ tends to $c_k'$ from the interior of the region $D_0$;
the function $F_m$ is from Lemma \ref{diff}.
\end{lemma}
{\bf Proof.}
Using Lemma \ref{diff}, we get
\begin{equation}\label{q1}
\frac{\partial}{\partial\l_m}\sum_{n=1}^{M}\sum_{l=1}^{r_n}\oint_{|\l^{(l)}-\l_n|=\rho}\phi_\l\phi\,d\l=
\sum_{l=1}^{r_m}\oint_{|\l^{(l)}-\l_m|=\rho}(\phi_\l^2+\phi\phi_{\l\l})d\l-
\end{equation}
\begin{equation*}
-\sum_{n=1}^{M}\sum_{l=1}^{r_n}\oint_{|\l^{(l)}-\l_n|=\rho}(F_m\phi_\l+[F_m]_\l)\phi_\l
+\phi([F_m]_\l\phi_\l+F_m\phi_{\l\l}+[F_m]_{\l\l})\,d\l=
\end{equation*}
\begin{equation*}
=-\sum_{l=1}^{r_m}\oint_{|\l^{(l)}-\l_m|=\rho}|\phi_\l|^2\,d\bar\l+
\sum_{n=1}^{M}\sum_{l=1}^{r_n}\oint_{|\l^{(l)}-\l_n|=\rho}F_m|\phi_\l|^2\,d\bar\l+
\phi_{\bar\l}[F_m]_\l\,d\bar\l\ \, .
\end{equation*}
For the integrals around the infinities we have the equality
\begin{equation}\label{infty12}
\frac{\partial}{\partial \l_m}\sum_{k=1}^N\oint_{|\l^{(k)}|=1/\rho}\phi_\l\phi\,d\l=
\sum_{k=1}^N\oint_{|\l^{(k)}|=1/\rho}F_m|\phi_\l|^2\,d\bar\l+\phi_{\bar\l}[F_m]_\l\,d\bar\l\ .
\end{equation}

Applying the Cauchy theorem to the (holomorphic) function $[F_m]_\l\phi_\l$, we get
\begin{equation}\label{c2}
\sum_{k=1}^g\int_{C_k\cup C_k^-}[F_m]_\l\phi_\l\,d\l=
-\left(\sum_{n=1}^{M}\sum_{l=1}^{r_n}\oint_{|\l^{(l)}-\l_n|=\rho}+\sum_{k=1}^N\oint_{|\l^{(k)}|=1/\rho}\right)
[F_m]_\l\phi_\l\,d\l\ .
\end{equation}
By (\ref{q1}), (\ref{infty12}) and (\ref{c2})
\begin{equation}\label{finT}
-\frac{i}{2}\frac{\partial}{\partial\l_m}\Big[T_\rho\Big]=
\frac{i}{2}\sum_{l=1}^{r_m}\oint_{|\l^{(l)}-\l_m|=\rho}|\phi_\l|^2\,d\bar\l-
\end{equation}
\begin{equation*}-\frac{i}{2}
\left\{\left(\sum_{n=1}^{M}\sum_{l=1}^{r_n}\oint_{|\l^{(l)}-\l_n|=\rho}+\sum_{k=1}^N\oint_{|\l^{(k)}|=1/\rho}\right)
\big(F_m|\phi_\l|^2d\bar\l-[F_m]_\l\phi_\l\,d\l+[F_m]_\l\phi_{\bar\l}\,d\bar\l \big)\right\}+
\end{equation*}
\begin{equation*}
+\frac{i}{2}\sum_{k=1}^g\int_{C_k\cup C_k^-}[F_m]_\l\phi_\l\,d\l\ .
\end{equation*}
Denote the expression in the large braces by $\Sigma_2$. We claim that
\begin{equation}\label{claim}
-\frac{i}{2}\Sigma_2=-\frac{i}{2}\left(\sum_{l=1}^{r_m}\oint_{|\l^{(l)}-\l_m|=\rho}|\phi_\l|^2\,d\bar\l
-2\pi i\sum_{n=1}^M\frac{1}{(r_n-1)!}
\left(1-\frac{1}{r_n^2}\right)\Phi_n^{(r_n)}(0)\right)+o(1)\;,
\end{equation}
where the function $\Phi_n$ is from Corollary \ref{kratnost1}.

To prove this we set
\ben
I_1^n(\rho)=\sum_{p=1}^{r_n}\oint_{|\lambda^{(p)}-\lambda_n|=\rho}F_m|\phi_\lambda|^2d \bar\lambda\;;
\;\;\;\;\ I_2^n(\rho)=\sum_{p=1}^{r_n}\oint_{|\lambda^{(p)}-\lambda_n|=\rho}
[F_m]_\lambda\phi_\lambda\,d\lambda\;;
\een
\ben
I_3^n(\rho)=\sum_{p=1}^{r_n}\oint_{|\lambda^{(p)}-\lambda_n|=
\rho}[F_m]_\lambda\phi_{\bar\lambda}d\bar\lambda\;.
\een

 By Corollary \ref{kratnost1} we have
\ben
I_1^n(\rho)=
\delta_{nm}\sum_{p=1}^{r_n}\oint_{|\lambda^{(p)}-\lambda_n|=\rho}|\phi_\lambda|^2
d\bar\lambda
\een
\ben
+\oint_{|x_n|=\rho^{1/r_n}}\left[\frac{1}{(r_n-1)!}\Phi_n^{(r_n-1)}(0)x_n^{r_n-1}+
\frac{1}{r_n!}\Phi_n^{(r_n)}(0)x_n^{r_n}+O(|x_n|^{r_n+1}\right]
\een
\ben
\times\left(\frac{|\phi_{x_n}^{int}|^2}{r_n x_n^{r_n-1}{\bar x}_n^{r_n-1}}+
\frac{1-r_n}{r_n^2}\frac{\phi_{x_n}^{int}}{{\bar x}_n^{r_n}x_n^{r_n-1}}
+\frac{1-r_n}{r_n^2}\frac{\phi_{{\bar x}_n}^{int}}{{\bar x}_n^{r_n-1}x_n^{r_n}}
+\left(\frac{1}{r_n}-1\right)^2\frac{1}{x_n^{r_n}{\bar x}_n^{r_n}}\right)r_n{\bar
x}_n^{r_n-1}d{\bar x}_n
\een
\ben
=\delta_{nm}\sum_{p=1}^{r_n}\oint_{|\lambda^{(p)}-\lambda_n|=\rho}|\phi_\lambda|^2
d\bar\lambda+2\pi i\frac{(1/r_n-1)^2}{(r_n-1)!}\Phi_n^{(r_n)}(0)+2\pi i\frac{1-r_n}{r_n(r_n-1)!}
\Phi_n^{(r_n-1)}(0)\phi_{x_n}^{int}(0)+o(1)
\een
as $\rho\to 0$.

We get also
\ben
I_2^n(\rho)=
-2\pi i\left(\frac{1}{r_n}-1\right)\frac{1}{(r_n-1)!}\Phi_n^{(r_n)}(0)-2\pi i\frac{1}{r_n(r_n-2)!}
\phi_{x_n}^{int}(0)\Phi_n^{(r_n-1)}(0)+o(1)
\een
and
\ben
I^n_3(\rho)=
2\pi i\frac{(1/r_n-1)}{(r_n-1)!}\Phi_n^{(r_n)}(0)+o(1)\;.
\een

We note that
\ben
I_1^n-I_2^n+I_3^n=\delta_{nm}\sum_{p=1}^{r_n}\oint_{|\lambda^{(p)}-\lambda_n|=\rho}|\phi_\lambda|^2
d\bar\lambda+\frac{2\pi
i}{(r_n-1)!}\Phi_n^{(r_n)}(0)[(1/r_n-1)^2+2(1/r_n-1)]+o(1)
\een
\ben
=\delta_{nm}\sum_{p=1}^{r_n}\oint_{|\lambda^{(p)}-\lambda_n|=\rho}|\phi_\lambda|^2
d\bar\lambda-\frac{2\pi
i}{(r_n-1)!}\left(1-\frac{1}{r_n^2}\right)\Phi_n^{(r_n)}(0)+o(1)\;.
\een
It is easy to  verify that
\ben
\sum_{k=1}^N\Big(\oint_{|\lambda^{(k)}|=1/\rho}F_m|\phi_\lambda|^2d\bar\lambda-
\oint_{|\lambda^{(k)}|=1/\rho}[F_m]_\lambda\phi_\lambda\,d\lambda+\oint_{|\lambda^{(k)}|=1/\rho}[F_m]_\lambda
\phi_{\bar\lambda}d\bar\lambda\Big)=o(1)\;,
\een
so we get (\ref{claim}).

The function $F_m(2\phi_{\l\l}-\phi_\l^2)$ is holomorphic outside of the ramification points, the infinities and the cuts.
Applying to it the Cauchy theorem and making use of Lemma \ref{shwartz} and the asymptotics from Lemma \ref{funR},
we get the equality
\begin{equation}\label{Cauchy}
2\pi i\sum_{n=1}^M\frac{1}{(r_n-1)!}
\left(1-\frac{1}{r_n^2}\right)\Phi_n^{(r_n)}(0)=
\end{equation}
\begin{equation*}
=-\frac{4\pi i}{(r_m-2)!\,r_m}\left(\frac{d}{dx_m}\right)^{r_m-2}R^{\omega, x_m}(x_m)\Big|_{x_m=0}
+
\sum_{k=1}^g
\int_{C_k\cup C_k^-}\left\{F_m(2\phi_{\l\l}-\phi_\l^2)\,\right\}d\l.
\end{equation*}
Summarizing (\ref{finT}), (\ref{claim}) and (\ref{Cauchy}), we get (\ref{stch}).

$\square$

Now we shall differentiate with respect to $\l_m$ the remaining terms in (\ref{str2}).
Denote by $L_{k; m}$, $\O_{; m}$ the derivatives $\frac{\partial}{\partial \l_m}L_k$, $\frac{\partial}
{\partial\l_m}\O$.
Since $\phi_\l$ is holomorphic with respect to $\l_m$, we have $[\phi_{\bar\l}]_{\l_m}=0$.
Thus,
\begin{equation}\label{ostal}
\frac{\partial}{\partial \l_m}\left[
-\frac{i}{2}\sum_{k=2}^g\int_{C_k}
\phi_{\bar\l}(\l, \bar\l)\log|L_k'(\Omega(\l))|^2\,d\bar\l-\frac{i}{2}\sum_{k=2}^g\int_{C_k}\phi(\l, \bar\l)
\frac{L_k''(\Omega(\l))}{L_k'(\Omega(\l))}\Omega'(\l)\,d\l\right]=
\end{equation}
\begin{equation*}
=\frac{i}{2}\sum_{k=2}^g\int_{C_k}\phi_\l\frac{L_{k; m}'(\Omega(\l))+L_k''(\Omega(\l))\Omega_{; m}(\l)}
{L_k'(\Omega(\l))}\,d\l+\frac{i}{2}\sum_{k=2}^g
(F_m\phi_\l+[F_m]_\l)\frac{L_k''(\Omega(\l))}{L_k'(\Omega(\l))}\Omega'(\l)\,d\l.
\end{equation*}
(We have used the equality
$$\phi_{\bar\l}\frac{\partial}{\partial\l_m}\log|L_k'(\Omega(\l))|^2\,d\bar\l+
\phi\frac{\partial^2}{\partial\l \partial\l_m}\log|L_k'(\Omega(\l))|^2\,d\l=$$
$$=
d\big(\phi\frac{\partial}{\partial\l_m}\log|L_k'(\Omega(\l))|^2\big)-\phi_\l
\frac{\partial}{\partial\l_m}\log|L_k'(\Omega(\l))|^2\,d\l$$
and  Lemma \ref{diff}.)

To finish the proof  we have to rewrite the last term at the right hand side of (\ref{stch})
as follows
\begin{equation}\label{starchl}
\frac{i}{2}\int_{C_k\cup C_k^-}\left\{F_m(2\phi_{\l\l}-\phi_\l^2)+[F_m]_\l\phi_\l\right\}\,d\l
=\frac{i}{2}\int_{C_k\cup C_k^-}\phi_\l\phi_{\l_m}\,d\l=
\end{equation}
\begin{equation*}=
\frac{i}{2}\int_{C_k}\phi_\l\phi_{\l_m}-\Big(\phi_\l+
\frac{L_k''(\Omega(\l))}{L_k'(\Omega(\l))}\Omega'(\l)\Big)\Big(\phi_{\l_m}+
\frac{L_{k; m}'(\Omega(\l))+L_k''(\Omega(\l))\Omega_{; m}(\l)}{L_k'(\Omega(\l))}\Big)\,d\l=
\end{equation*}
\begin{equation*}
=-\frac{i}{2}\int_{C_k}\Big[\phi_\l\frac{L_{k; m}'(\Omega(\l))+L_k''(\Omega(\l))\Omega_{; m}(\l)}{L_k'(\Omega(\l))}
+\phi_{\l_m}\frac{L_k''(\Omega(\l))}{L_k'(\Omega(\l))}\Omega'(\l)+
\end{equation*}
\begin{equation*}+
\frac{L_k''(\Omega(\l))}{L_k'(\Omega(\l))}\Omega'(\l)
\frac{L_{k; m}'(\Omega(\l))+L_k''(\Omega(\l))\Omega_{; m}(\l)}{L_k'(\Omega(\l))}\Big]\,d\l.
\end{equation*}

Collecting (\ref{str2}), (\ref{stch}), (\ref{ostal}) and (\ref{starchl})
and using the equality
$$\phi_{\l_m}=\frac{\Omega_{; m}'(\l)}{\Omega'(\l)},$$
 we get
\begin{equation}\label{FFin}
\frac{\partial S_\rho}{\partial \l_m}+o(1)=\frac{2\pi}{(r_m-2)!\,r_m}\left
(\frac{d}{dx_m}\right)^{r_m-2}R^{\omega, x_m}\Big|_{x_m=0}
-\frac{i}{2}\sum_{k=2}^g\int_{C_k}\frac{L_k''(\Omega(\l))L_{k; m}'(\Omega(\l))}{\big[L_k'(\Omega(\l))\big]^2}
\Omega'(\l)\,d\l-
\end{equation}
\begin{equation*}
-\frac{i}{2}\sum_{k=2}^g\int_{C_k}\left[\frac{L_k''(\Omega(\l))}{L_k'(\Omega(\l))}\right]^2
\Omega'(\l)\Omega_{; m}(\l)\,d\l-i\sum_{k=2}^g\int_{C_k}
\frac{L_k''(\Omega(\l))}{L_k'(\Omega(\l))}\Omega_{; m}'(\l)\,d\l.
\end{equation*}
Since $\{L_k(\omega), \omega\}\equiv 0$, the last two terms in (\ref{FFin}) cancel (one should
beforehand integrate the last term by parts).
For the second term we have the equality (\cite{ZT3}):
$$-\frac{i}{2}\int_{C_k}\frac{L_k''(\Omega(\l))L_{k; m}'(\Omega(\l))}{\big[L_k'(\Omega(\l))\big]^2}
\Omega'(\l)\,d\l
=-4\pi\frac{l_{k; m}}{l_k}\ \,.$$
To prove Theorem \ref{sch} it is sufficient to observe that
the term $o(1)$ in (\ref{FFin})
is uniform with respect to parameters $(\lambda_1, \dots, \lambda_M)$ belonging to a compact neighborhood of the initial
point $(\lambda_1^0, \dots, \lambda_M^0)$.

$\square$

\subsection{The Liouville action and the Fuchsian uniformization}
\subsubsection{The metric of constant curvature $-1$ on $\L$ and its dependence upon the branch points}

 The covering $\L$ is biholomorphically equivalent to the quotient space ${\mathbb H}/\Gamma$, where
 ${\mathbb H}=\{z\in {\mathbb C}\ :\ \Im z>0\}$, $\Gamma$ is a strictly hyperbolic Fuchsian group.
 Denote by $\pi_\Gamma: {\mathbb H}\rightarrow \L$ the natural projection.
Let $x$ be a local parameter on $\L$, introduce the metric $e^{\chi(x, \bar x)}|dx|^2$ of the constant
curvature $-1$ on $\L$ by the equality
\begin{equation}\label{locmetr}
e^{\chi(x, \bar x)}|dx|^2=\frac{|dz|^2}{|\Im z|^2},
\end{equation}
where $z\in {\mathbb H}$, $\pi_\Gamma(z)=x$.
As usually we specify the functions $\chi^{ext}(\l, \bar\l)$, $\chi^{int}(x_m, \bar x_m)$, $m=1, \dots, M$ and
$\chi^{\infty}(\zeta_k, \bar\zeta_k)$, $k=1, \dots, N$ setting $x=\l$, $x=x_m$
and $x=\zeta_k$ in (\ref{locmetr}).

Set $R^{z, x}=\{z, x\}$, where $z\in {\mathbb H}$, $\pi_\Gamma(z)=x$.
Clearly, Lemmas \ref{as} and \ref{shwartz} still stand with $\chi^{ext}$, $R^{z, x}$ instead of
$\phi^{ext}$ and $R^{\omega, x}$, whereas Lemma \ref{diff} should be reconsidered, since the Fuchsian
uniformization map depends upon the branch points nonholomorphically.

Introduce the metric $e^{\psi(\omega, \bar\omega)}|d\omega|^2$
 of constant curvature $-1$ on $D_0$ (see the previous section) by
the equation
$$e^{\psi(\omega, \bar\omega)}|d\omega|^2=\frac{|dz|^2}{|\Im z|^2},$$
where $\pi_\Sigma(\omega)=\pi_\Gamma(z)$.
Then there is the following relation between the derivatives of the function $\psi$:
\begin{equation}\label{zota}
\psi_{\l_m}(\omega, \bar\omega)+\psi_\omega(\omega, \bar\omega){\mathbb F}_m(\omega, \bar\omega)+
[{\mathbb F}_m]_\omega(\omega, \bar\omega)=0,
\end{equation}
where ${\mathbb F}$ is a continuously differentiable function on $D_0$;
(the proof of (\ref{zota}) is parallel to the one in (\cite{ZT3})).

We shall now prove the analog of  (\ref{zota}) and Lemma \ref{diff} for the function $\chi=\chi^{ext}$.
\begin{lemma}\label{analog}
There is the following relation between the derivatives of the function $\chi$:
\begin{equation}\label{ana}
\frac{\partial \chi(\l, \bar\l)}{\partial \l_m}+{\mathcal F}_m(\l, \bar\l)\frac{\partial \chi(\l, \bar\l)}{\partial \l}
+\frac{\partial{\mathcal F}_m(\l, \bar\l)}{\partial \l}=0,
\end{equation}
where
\begin{equation}\label{Fkras}
{\mathcal F}_m(\l, \bar\l)={\mathbb F}_m(\Omega_0(\l), \overline{\Omega_0(\l)})\frac{1}{\Omega_0'(\l)}+
F_m(\l).
\end{equation}
Here $F_m=-\frac{[\Omega_0]_{\l_m}}{[\Omega_0]_\l}$ is the function from Lemma \ref{diff}, ${\mathbb F}_m$ is
the function from (\ref{zota}).
\end{lemma}

{\bf Proof.} Since
$$e^{\chi(\l, \bar\l)}|d\l|^2=e^{\psi(\Omega_0(\l), \overline{\Omega_0(\l)})}|\Omega_0'(\l)|^2|d\l|^2,$$
we have the equality
\begin{equation}\label{summa}
\chi(\l, \bar\l)=\psi(\Omega_0(\l), \overline{\Omega_0(\l)})+\phi(\l,\bar\l),
\end{equation}
where $\phi(\l, \bar\l)=\log |\Omega_0'(\l)|^2$ is the function from (\ref{defmetric}).
Differentiating (\ref{summa}) with respect to $\l_m$ via formulas (\ref{zota}) and (\ref{QuaziAlfors1}),
after some easy calculations we get (\ref{ana}).

$\square$

\begin{remark}\label{skachok}{\rm Observe that the function ${\mathcal F}_m$ does not have jumps at the cycles $C_k$,
whereas the both terms at the right hand side of (\ref{Fkras}) do.
This immediately follows from the formulas
$$F^-_m(\l)=F^+_m(\l)-\frac{L_{k; m}(\Omega_0^+(\l))}{L_k'(\Omega_0^+(\l))[\Omega_0^+]_\l(\l)},$$
$$[\Omega^-_0]_\l(\l)=L_k'(\Omega^+_0(\l))[\Omega^+_0]_\l(\l)$$
and the formula from \cite{ZT3}:
$${\mathbb F}_m\circ L_k={\mathbb F}_mL_k'+L_{k; m}.$$
Here the indices $+$ and $-$ denote the limit values of the corresponding functions at the
"$c_k$" and the "$c_k'$"
sides of the cycle $C_k$.}
\end{remark}
\begin{lemma}\label{delta}
Fix a number $m=1, \dots, M$. Then for any $n=1, \dots, M$
the following asymptotics holds
\begin{equation}\label{point-m}
{\mathcal F}_m(\l_n+x_n^{r_n}, \bar \l_n+{\bar x}_n^{r_n})=\delta_{mn}+a_nx_n^{r_n-1}
+b_n{\bar x}_nx_n^{r_n-1}+c_nx_n^{r_n}+O(|x_n|^{r_n+1})
\end{equation}
as $x_m\rightarrow 0$; here $a_n, b_n, c_n$ are some complex constants.

At the infinity of the $k$-th sheet of the covering $\L$ there is the asymptotics
\begin{equation}\label{besk-k}
{\mathcal F}_m(\l, \bar\l)=A_k\l^2+B_k\l+C_k\l^2\bar\l^{-1}+O(1)
\end{equation}
as $\l\rightarrow\infty^{(k)}$; here $\infty^{(k)}$ is the point at infinity of the $k$-th sheet of the covering
$\L$; $A_k, B_k, C_k$ are some complex constants.
\end{lemma}

{\bf Proof.} This follows from Corollary \ref{kratnost1}, asymptotics (\ref{d1}) and formula (\ref{Fkras}).

$\square$

\subsubsection{The regularized Liouville action}

Here we define the regularized integral
$${\rm reg}\int_\L(|\chi_\l|^2+e^\chi) \,dS$$
and calculate its derivatives with respect to the branch points $\l_m$.

Set $\Lambda_\rho^k=\{\l\in \L^{(k)}\ : \ \forall m\ \ \ |\l-\l_m|>\rho\  \& \ |\l|<1/\rho\}$,
where $P_m$ are all the ramification points which belong to the $k$-th sheet $\L^{(k)}$ of the covering $\L$.
To the sheet $\L^{(k)}$ there corresponds the function $\chi_k^{ext}: \L^{(k)}\rightarrow {\mathbb R}$
which is smooth in any domain $\Lambda_\rho^k$, $\rho>0$.

The function $\chi_k^{ext}$ has finite limits at the cuts (except the endpoints which are the ramification points);
at the ramification points and at the infinity it possesses the same asymptotics as the function $\phi_k^{ext}$
from the previous section.

Observe also that the function $e^{\chi_k^{ext}}$ is integrable on $\L^{(k)}$.
Set
\begin{equation}\label{qu-qu}
T_\rho=\sum_{k=1}^N\int_{\Lambda_\rho^k}|\partial_\l\chi_k^{ext}|^2\,dS.
\end{equation}
Then there exists the finite limit
\begin{equation}\label{reg2}
{\rm reg}\int_\L(|\chi_\l|^2+e^\chi)\,dS=\lim_{\rho\to 0}
\left(T_\rho+\sum_{k=1}^N\int_{\L^{(k)}}e^{\chi_k^{ext}}\,dS
+(4N+\sum_{m=1}^M\frac{(r_m-1)^2}{r_m})2\pi\log \rho\right).
\end{equation}
Set
\begin{equation}\label{Main2}
{\mathbb S}_{Fuchs}(\l_1, \dots, \l_M)=\frac{1}{2\pi}{\rm reg}\int_\L(|\chi_\l|^2+e^\chi)\,dS+2
\sum_{n=1}^M(r_n-1)\chi^{int}(x_n)\Big|_{x_n=0}-2\sum_{k=1}^N\chi^\infty(\zeta_k)\Big|_{\zeta_k=0}.
\end{equation}
Now we state the main result of this section.
\begin{theorem}\label{M-fuchs}
For any $m=1, \dots, M$ the following equality holds
\begin{equation}\label{tfuchs}
\frac{\partial{\mathbb S}_{Fuchs}(\l_1, \dots, \l_M)}{\partial \l_m}=
\frac{1}{(r_m-2)!\,r_m}\left(\frac{d}{dx_m}\right)^{r_m-2}R^{z, x_m}\Big|_{x_m=0}.
\end{equation}
\end{theorem}

{\bf Proof.} Set  $\Lambda_\rho=\cup_{k=1}^N\Lambda_\rho^k$.
Then
\begin{equation}\label{t1}
\frac{\partial}{\partial \l_m}T_\rho=
\frac{i}{2}\sum_{k=1}^{r_m}\oint_{|\l^{(k)}-\l_m|=\rho}|\partial_\l\chi|^2\,d\bar\l+
\int_{\Lambda_\rho}\frac{\partial}{\partial\l_m}|\partial_\l\chi|^2\,dS.
\end{equation}
By (\ref{ana}) the last term in (\ref{t1}) can be rewritten as
\begin{equation}\label{t2}
\int_{\Lambda_\rho}\frac{\partial}{\partial\l_m}|\partial_\l\chi|^2\,dS=
\end{equation}
\begin{equation*}
\int_{\Lambda_\rho}\Big(((2\chi_{\l\l}-\chi_\l^2)[{\mathcal F}_m])_{\bar\l}
-2(\chi_\l[{\mathcal F}_m]_{\bar\l})_\l+(\chi_\l[{\mathcal F}_m]_\l)_{\bar\l}-
(\chi_{\bar\l}[{\mathcal F}_m]_\l)_\l-(|\chi_\l|^2[{\mathcal F}_m])_\l\Big)\,dS=
\end{equation*}
\begin{equation*}
=-\frac{i}{2}\int_{\partial \Lambda
_\rho}(2\chi_{\l\l}-\chi_\l^2){\mathcal F}_m\,d\l+2\chi_\l[{\mathcal F}_m]_{\bar\l}\,d\bar\l+
\chi_\l[{\mathcal F}_m]_\l\,d\l+\chi_{\bar\l}[{\mathcal F}_m]_\l\,d\bar\l+|\chi_\l|^2{\mathcal F}_m\,d\bar\l=
\end{equation*}
\begin{equation*}
-\frac{i}{2}\sum_{n=1}^M\Big[I_1^n+2I_2^n+I_3^n+I_4^n+I_5^n\Big]-
\frac{i}{2}\sum_{k=1}^N\Big[J_1^{\infty, k}+J_2^{\infty, k}+J_3^{\infty, k}+J_4^{\infty, k}+J_5^{\infty, k}\Big],
\end{equation*}
where
$$I_1^n=\sum_{l=1}^{r_n}\oint_{|\l^{(l)}-\l_n|=\rho}(2\chi_{\l\l}-\chi_\l^2){\mathcal F}_m\,d\l,$$
$$J_1^{\infty, k}=\oint_{|\l^{(k)}|=1/\rho}(2\chi_{\l\l}-\chi_\l^2){\mathcal F}_m\,d\l$$
and
the terms $I_p^n$ and $J_p^{\infty, k}$, $p=2, 3, 4, 5$ are the similar sums of integrals and  integrals
with integrands $\chi_\l[{\mathcal F}_m]_{\bar\l}\,d\bar\l$,
$\chi_\l[{\mathcal F}_m]_\l\,d\l$,
$\chi_{\bar\l}[{\mathcal F}_m]_\l\,d\bar\l$
and $|\chi_\l|^2{\mathcal F}_m\,d\bar\l$ respectively.
It should be noted that the circles $|\l-\l_n|=\rho$ are clockwise oriented whereas the circles
$|\l|=1/\rho$ are counter-clockwise oriented.
Using (\ref{point-m}), we get
\begin{equation}\label{t3}
I_1^n=
\oint_{|x_n|=\rho^{1/r_n}}\Big[\frac{2R^{z, x_n}(x_n)}{r_nx_n^{2r_n-2}}+\left(1-\frac{1}{r_n^2}\right)
\frac{1}{x_n^{2r_n}}\Big]\Big(\delta_{mn}+a_nx_n^{r_n-1}+b_n{\bar x}_nx_n^{r_n-1}+
\end{equation}
\begin{equation*}
+c_nx_n^{r_n}+O(|x_n|^{r_n+1})
\Big)r_nx_n^{r_n-1}\,dx_n=-\delta_{nm}\frac{4\pi i}{(r_n-2)!\,r_n}\left(\frac{d}{dx_n}\right)^{r_n-2}R^{z,
 x_m}(0)-
\end{equation*}
\begin{equation*}
-2\pi ir_n\left(1-\frac{1}{r_n^2}\right)c_n+o(1).
\end{equation*}
In the same manner we get
\begin{equation}\label{t4}
I_2^n=o(1), \ \ I_3^n=-2\pi i\left(\frac{r_n-1}{r_n}a_n\chi^{int}_{x_n}(0)+r_n\big(\frac{1}{r_n}-1\big)c_n
\right)+
o(1),
\end{equation}
and
\begin{equation}\label{t5}
I_4^n=2\pi i\left(\frac{1}{r_n}-1\right)r_nc_n+o(1),\ \ \ I_5^n=\delta_{mn}\sum_{l=1}^{r_n}\oint_{
|\l^{(l)}-\l|=\rho}|\chi_\l|^2\,d\bar\l+
\end{equation}
\begin{equation*}
+2\pi i\chi^{int}_{x_n}(0)\frac{1-r_n}{r_n}a_n+
2\pi i\left(\frac{1}{r_n}-1\right)^2r_nc_n+o(1).
\end{equation*}
Using (\ref{besk-k}), we get also
\begin{equation}\label{t6}
J_1^{\infty, k}=o(1),\ \ \ J_2^{\infty, k}=o(1), \ \ \ J_3^{\infty, k}=-4\pi i(A_k\chi_{\zeta_k}^\infty(0)+B_k)+o(1),
\end{equation}
and
\begin{equation}\label{t7}
J_4^{\infty, k}=4\pi i B_k+o(1),\ \ \ J_5^{\infty, k}=-4\pi i(A_k\chi_{\zeta_k}^\infty(0)+2B_k)+o(1).
\end{equation}
Summarizing (\ref{t1}-\ref{t7}), we have
\begin{equation}\label{promez}
\frac{\partial}{\partial \l_m}T_\rho=\frac{2\pi}{(r_m-2)!\,r_m}\left(\frac{d}{dx_m}\right)^{r_m-2}
R^{z, x_m}(0)+2\pi\sum_{n=1}^M\frac{1-r_n}{r_n}\left(a_n\chi^{int}_{x_n}(0)+c_n\right)-
\end{equation}
\begin{equation*}
-4\pi\sum_{k=1}^N
\left(A_k\chi^\infty_{\zeta_k}(0)+B_k\right)+o(1).
\end{equation*}
To finish the proof we need the following lemma.
\begin{lemma}\label{svyaz}
The equalities hold
\begin{equation}\label{e1}
\frac{\partial}{\partial\l_m}\chi^{int}(x_n)\Big|_{x_n=0}=-\frac{1}{r_n}(a_n\chi^{int}_{x_n}(0)+c_n)
\end{equation}
and
\begin{equation}\label{e2}
\frac{\partial}{\partial\l_m}\chi^\infty(\zeta_k)\Big|_{\zeta_k=0}=A_k\chi^\infty_{\zeta_k}(0)+B_k.
\end{equation}
\end{lemma}
{\bf Proof.} We shall prove (\ref{e1}); (\ref{e2}) can be proved analogously.
Since
$$e^{\chi^{int}(x_n, {\bar x}_n)}|dx_n|^2=e^{\chi^{ext}(\l, \bar\l)}|d\l|^2,$$
we get
\begin{equation}\label{le-1}
\chi^{int}(x_n, {\bar x}_n)=\chi^{ext}(\l, \bar\l)-\left(\frac{1}{r_n}-1\right)\frac{1}{r_n^2}\log|\l-\l_n|^2
\end{equation}
and
\begin{equation}\label{le-2}
\chi^{ext}_{\l_m}(\l, \bar\l)=\chi^{int}_{\l_m}(x_n, {\bar x}_n)+{\rm const}\,\delta_{mn}\frac{1}{x_n^{r_n}}.
\end{equation}
By (\ref{point-m}) and (\ref{ana}) we have
\begin{equation}\label{le-3}
\chi^{ext}_{\l_m}(\l, \bar\l)=
\end{equation}
\begin{equation*}
=-\Big(\delta_{mn}+a_nx_n^{r_n-1}+b_{n}{\bar x}_nx_n^{r_n-1}+c_nx_n^{r_n}+O(|x_n|^{r_n+1})\Big)
\Big[\frac{1}{r_nx_n^{r_n-1}}\chi^{int}_{x_n}(x_n, \bar x_n)+\left(\frac{1}{r_n}-1\right)\frac{1}{x_n^{r_n}}\Big]
-
\end{equation*}
\begin{equation*}
-\frac{r_n-1}{r_n}a_n\frac{1}{x_n}-\frac{r_n-1}{r_n}b_n\frac{\bar x_n}{x_n}-c_n+O(|x_n|).
\end{equation*}
Now substituting (\ref{le-3}) in (\ref{le-2}) and comparing the coefficients near the zero power of
$x_n$, we get (\ref{e1}).

$\square$

Observe that
$$\frac{\partial}{\partial\l_m}\int_\L e^{\chi}\,dS=0$$
due to the Gauss-Bonnet theorem
and the term $o(1)$ in (\ref{promez}) is uniform with respect to $(\l_1, \dots, \l_M)$ belonging
to a compact neighborhood of the initial point $(\l_1^0, \dots, \l_M^0)$.
This together with (\ref{promez}) and Lemma \ref{svyaz} proves Theorem \ref{M-fuchs}.

$\square$

\begin{remark}\label{var}{\rm
Consider the functional defined by the right hand side of (\ref{Main2}). If we introduce variations
$\delta\chi$ which are smooth functions on $\L$ vanishing in neighborhoods of the branch points and the
infinities then the Euler-Lagrange equation for an extremal of this functional coinsides
with the Liouville equation
$$\chi_{\l\bar\l}=\frac{1}{2}e^\chi.$$
The last equation is equivalent to the condition that the metric $e^\chi|d\l|^2$ has constant curvature $-1$.
}
\end{remark}

\subsection{The modulus square of Bergmann and Wirtinger tau-functions in higher genus}

Now we are in a position to calculate the modulus square of Bergmann (and, therefore, Wirtinger)
tau-function. Actually, we shall give two equivalent answers:
one is given in terms of the Fuchsian uniformization of the surface $\L$
and the determinant of the Laplacian, another one
uses the Schottky uniformization and the holomorphic determinant of the Cauchy-Riemann operator
in the trivial line bundle over $\L$.

 Indeed, formula (\ref{detL}) and Theorem \ref{M-fuchs} imply the following statement.
\begin{theorem}\label{MF}
Let the regularized Liouville action ${\mathbb S}_{Fuchs}$ be given by formula (\ref{Main2}).
Then we have the following expression
for the modulus square $|\tau_B|^2$ of the Bergmann tau-function of the covering $\L$:
\begin{equation}\label{otvetFB}
|\tau_B|^2=e^{-{\mathbb S}_{Fuchs}/6}\frac{{\rm det}\Delta}{{\rm det}\,\Im\,{\mathbb B}}.
\end{equation}
For the modulus square $|\tau_W|^2$ of the Wirtinger tau-function we have the expression:
\begin{equation}\label{otvetFW}
|\tau_W|^2=e^{-{\mathbb S}_{Fuchs}/6}\frac{{\rm det}\Delta}{{\rm det}\,\Im\,{\mathbb B}}
\prod_{\beta\,{\rm  even}}\Big|\Theta[\beta](0\,|\,{\mathbb B})
\Big|^{-\frac{2}{4^{g-1}+2^{g-2}}}\;.
\end{equation}
\end{theorem}

  On the other hand, using formula (\ref{dbar}) and Theorem \ref{sch}, we get the following alternative
answer.
\begin{theorem}\label{MSCH} Let the regularized Dirichlet integral ${\mathbb S}_{Schottky}$ be given
by formula (\ref{Main}). Then the modulus square of the Bergmann and Wirtinger tau-functions
of the covering $\L$ can be expressed as follows:
\begin{equation}\label{otvetschB}
|\tau_B|^2=e^{-{\mathbb S}_{Schottky}/6}|{\rm det}\bar\partial|^2,
\end{equation}
\begin{equation}\label{otvetschW}
|\tau_W|^2=e^{-{\mathbb S}_{Schottky}/6}|{\rm det}\bar\partial|^2
\prod_{\beta\,{\rm  even}}\Big|\Theta[\beta](0\,|\,{\mathbb B})
\Big|^{-\frac{2}{4^{g-1}+2^{g-2}}}\;.
\end{equation}
\end{theorem}

\begin{remark}
{\rm Comparing (\ref{otvetschB}), (\ref{otvetFB}) and formula (3.3) for $|{\text det}\bar\partial|^2$
from (\cite{Zograf}),
we get the equality $${\mathbb S}_{Schottky}-{\mathbb S}_{Fuchs}=\frac{1}{2\pi}S,$$ where
$S$ is the Liouville action from \cite{ZT3}. Whether it is possible to prove this relation directly is an
open question.}
\end{remark}

\begin{remark}{\rm Looking at the formulas for the tau-functions
in genera $0$ and $1$ (and for genus 2 two-fold coverings), one may
believe that the expressions for the tau-functions
in higher genus can be also given in pure holomorphic terms, without any use of the Dirichlet integrals
and, especially, the Fuchsian uniformization.
At the least, the Dirichlet integral should be eliminated from the proofs in genus $0$ and $1$.
}
\end{remark}

\begin{remark}
{\rm
The Hurwitz space $H_{g, N}(1, \dots, 1)$ covers the Stein manifold
${\mathbb C}^{(M)}\setminus \Delta$ and, therefore, also is a Stein manifold.
Here $M=2g+2N-2$,
${\mathbb C}^{(M)}$ is the $M$-th symmetric power of ${\mathbb C}$, $\Delta=
\cup_{i, j}\{\l_i=\l_j\}$; the number of sheets of the covering
$$H_{g, N}(1, \dots, 1)\rightarrow{\mathbb C}^{(M)}\setminus \Delta$$
(or, equivalently, the degree of the Lyashko-Looijenga map)
is finite and equals (up to the factor $N!$) to the Hurwitz number $h_{g, N}$.
Due to Remark \ref{stein},
in case $g=0, 1$ the $12$-th power $\tau_W^{12}$ of the Wirtinger tau-function gives a global holomorphic
function on $H_{g, N}(1, \dots, 1)$.
It is interesting whether the $\tau_W^{12}$ has critical points.
It should be noted that for two-fold genus 2 coverings
the answer to the analogous question is negative. For such coverings
the $20$-th power of $\tau_W$  is just the Vandermonde determinant $V(\l_1, \dots, \l_6)$.
Applying to $V(\l_1, \dots, \l_6)$ the Euler theorem for homogeneous functions,
we see that the first derivatives of $\tau_W^{20}$ never vanish simultaneously
on $H_{2, 2}(1, 1)={\mathbb C}^{(6)}\setminus \Delta$.

It could also be very interesting to connect  the Wirtinger
tau-function with the Hurwitz numbers $h_{g, N}$ and elucidate the
relationship of our tau-functions with $G$-function of Frobenius manifolds \cite{Strachan}.
}
\end{remark}

\end{document}